\documentclass[apj,twocolumn]{emulatearxiv}
\usepackage{amsmath}

\begin{document}

\title{Hydrodynamic Simulations of Pre-Supernova Outbursts in Red Supergiants: Asphericity and Mass Loss}


\author{Shing-Chi Leung\thanks{Email address: shingchi.leung@ipmu.jp}}

\affiliation{TAPIR, Walter Burke Institute for Theoretical Physics, 
Mailcode 350-17, Caltech, Pasadena, CA 91125, USA} 
 
\author{Jim Fuller\thanks{Email address: shingchi.leung@ipmu.jp}}

\affiliation{TAPIR, Walter Burke Institute for Theoretical Physics, 
Mailcode 350-17, Caltech, Pasadena, CA 91125, USA} 

\date{\today}

\begin{abstract}


The activity of massive stars approaching core-collapse can strongly affect the appearance of the star and its subsequent supernova. Late-phase convective nuclear burning generates waves that propagate toward the stellar surface, heating the envelope and potentially triggering mass loss. In this work, we improve on previous one-dimensional models by performing two-dimensional simulations of the pre-supernova mass ejection phase due wave heat deposition. Beginning with stellar evolutionary models of a 15 $M_{\odot}$ red supergiant star during core O-burning, we treat the energy deposition rate and duration as model parameters and examine the mass-loss dependence and the pre-explosion morphology accordingly. Unlike one-dimensional models, density inversions due to wave heating are smoothed by Rayleigh-Taylor instabilities, and the primary effect of wave heating is to radially expand the star's hydrogen envelope. For low heating rates with long durations, the expansion is nearly homologous, whereas high but short-lived heating can generate a shock that drives envelope expansion and results in a qualitatively different density profile at the time of core-collapse. Asymmetries are fairly small, and large amounts of mass loss are unlikely unless the wave heating exceeds expectations.
We discuss implications for pre-supernova stellar variability and supernovae light curves.

\end{abstract}




\section{Introduction}
\label{sec:intro}

\subsection{Pre-Supernova Stellar Evolution}

Type II supernovae (SNe) are the explosions of massive hydrogen-rich stars, and it is widely accepted that the progenitors of type II-P SNe are red supergiants.  Observations indicate that progenitor stars of interacting Type IIn SNe usually produce pre-SN outbursts \citep{Ofek2014}, specific 
examples including SN 2009ip \citep{Mauerhan2013}, 
SN 2010mc \citep{Ofek2013} and 
SN 2015bh \citep{ElisaRosa2016,Ofek2016}. However, even fairly typical type IIP SNe show evidence for circumstellar material (CSM) that may arise from pre-SN mass loss \citep{khazov:16,yaron:17}.
Pre-SN mass eruption events have also been observed or hinted in other types of SNe, including Type IIb such as SN 2013cu \citep{GalYam2014,Groh2014,Graefener2016}, Type Ibn such as SN 2006jc \citep{Pastorello2007}, 
SN 2015G \citep{Shivvers2017}, SN 2015U \citep{Shivvers2016}, and even broad-lined Type Ic such as PTF11qcj \citep{Corsi2014} and 2018gep \citep{ho:19}.

Pre-SN mass ejection has been proposed to explain many features of both ordinary and super-luminous SNe. The interaction between the ejecta from the final core-collapse explosion and the previously ejected circumstellar material allows the efficient conversion of the ejecta kinetic energy to thermal energy by shock heating. A classical example is pulsation pair-instability SNe \citep{Woosley2007,Yoshida2016,Woosley2018,Leung2019PPISN}. Shock waves driven by unstable oxygen burning eject large amounts of mass prior to the star's collapse, creating dense CSM that allows for a super-luminous SN \citep{Sorokina2016,Morozova2017,Tolstov2017}.

For less massive stars, wave energy transport within the progenitor may be able to trigger smaller amounts of pre-SN mass loss \cite[see e.g.][]{Quataert2012}.
Due to the large energy generation rate by nuclear reactions, convective motions near the star's core become very energetic during advanced burning phases such as carbon, neon, oxygen, and silicon burning \cite[e.g.][]{Woosley2002}. 
The convective motion generates gravity waves that carry a power $L_{\rm wave}$ that
scales with the convective luminosity $L_{\rm con}$ as 
$L_{\rm wave} \sim M_{\rm con} L_{\rm con}$ \citep{Goldreich1990,Alvan2015}, where 
$M_{\rm con}$ is the mean Mach number of the convective motion.
The gravity waves propagate outwards and are partially reflected
by overlying convective shells, though some wave energy is still 
transferred outwards into acoustic waves after tunneling through these evanescent regions
(see, e.g., \citealt{Fuller2015,Takata2016} for applications in low-mass red-giant stars).

The acoustic waves dissipate via weak shock formation when the energy transport rate by wave exceeds the 
maximum possible wave flux \citep{Ro2017}, which drops rapidly at the transition from the helium shell to the hydrogen envelope. The thermalized wave energy, depending on how fast it is injected, can trigger expansion of the envelope and also mass ejection \citep{Fuller2017}. Recent studies of observational data have shown that pre-SN activity may be important for explaining the light curve shape \citep{Moriya2017,Morozova2017,moriya:18,Morozova2018}, though too much wave heat (or heat deposited over too long of a duration) is inconsistent with typical type II-P SNe \citep{Ouchi2019}. Instead, more impulsive wave heating is favored based on light curve modeling of SN 2017eaw, confirming that such energy
deposition can better reproduce SN light curves \citep{Morozova2019}. 

We note that wave-driven outbursts may not occur in many SN progenitors, as demonstrated by SN progenitor monitoring \citep{kochanek:17,johnson:18} that rules out luminous and long-lasting outbursts in several nearby type II-P SNe. The purpose of this paper is not to determine the frequency or likelihood of wave-driven outbursts, but rather their effect on the progenitor's structure if they do occur.

\subsection{Motivation}

\cite{Fuller2017} and \cite{Fuller2018} compute one-dimensional stellar evolutionary models for a $15 ~M_{\odot}$ star model with and without a hydrogen envelope.
They demonstrate that wave heat has very different effects in hydrogen-rich and hydrogen-poor stars. In the latter, the energy can eject $\sim 10^{-2}-10^{-1} \, M_\odot$ in an optically thick super-Eddington wind from the surface of the star. In hydrogen-rich stars, however, wave energy is thermalized at the base of the hydrogen rich envelope and is likely insufficient to unbind the entire envelope. While small amounts of material may be unbound by a weak shock excited by wave heating, the one-dimensional models exhibit an inflation of the envelope as the extra thermal pressure blows a low-density ``bubble" at the base of the hydrogen envelope.  

However, the energy deposition can give rise to non-radial hydrodynamical instabilities that cannot be captured in any of the previous one-dimensional simulations. As mentioned above, wave energy deposition in a low-density bubble causes it to have a high thermal pressure but a low density. As it expands into the low-pressure and high-density envelope, Rayleigh-Taylor instability will occur at the interface between the bubble and the overlying envelope, strongly modifying the subsequent dynamics of the envelope's response. It therefore becomes interesting to examine multi-dimensional simulations of the star's response, which have never been performed in a systematic way for this process. In this work, we explore the general behaviour of the envelope in response to such energy deposition and examine the aspherical effects on the pre-SN mass loss and the envelope structure.


The organization of the paper is as follows. In Section \ref{sec:methods}, 
we describe the 
hydrodynamical formalism for modeling the wave energy deposition
in the stellar envelope and how we construct the initial model. 
Then in Section \ref{sec:results} we present a benchmark model which demonstrates the typical effects of wave heating in multiple dimensions.
In Section \ref{sec:massloss}, we 
examine how the mass loss process and the 
envelope structure changes with the energy deposition
and its duration. In Section \ref{sec:discussions}, we discuss
implications for supernovae and progenitor stars with different wave heating rates and durations.
In the Appendix, we further show how our simulations depend
some of the input parameters, including the mesh resolution,
the initial perturbation and the boundaries.

\section{Methods}
\label{sec:methods}

\subsection{Stellar Model}

\begin{figure*}
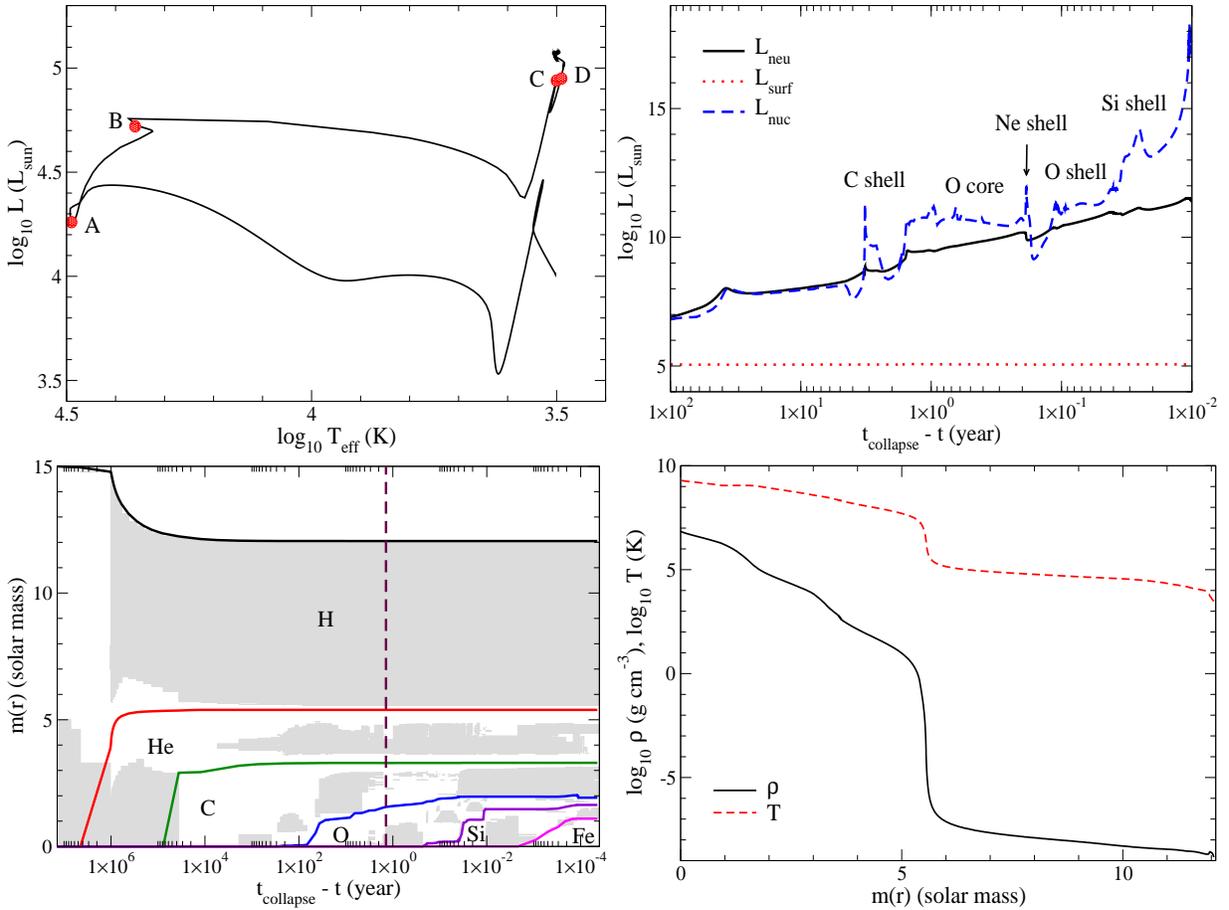

\centering
\includegraphics*[width=8cm,height=6cm]{HR_diagram.eps}
\includegraphics*[width=8cm,height=6cm]{Lnuc_time_plot.eps}
\includegraphics*[width=8cm,height=6cm]{kippenhahn_diag.eps}
\includegraphics*[width=8cm,height=6cm]{initial_profile84.eps}
\caption{
{\bf Top left:} HR diagram of our 15 $M_{\odot}$ stellar model. The letters A - D stand for the onset and end of H-burning (A and B), and He-burning (C and D) respectively. 
{\bf Top right:} Nuclear, neutrino and surface luminosities of the star against time. 
{\bf Bottom left:} Kippenhahn diagram of the model. Colored lines are the mass coordinates of the He, C, O, Si, and Fe cores. Vertical dashed line indicates the moment where the snapshot is taken for the hydrodynamics simulation. 
{\bf Bottom right:} Density and temperature profiles at the onset of the O-burning phase.
}
\label{fig:evol}
\end{figure*}

To construct the initial models for the hydrodynamics run,
we first use the one-dimensional stellar evolutionary code
MESA (Modules for the Experiments in Stellar Astrophysics, \citealt{Paxton2011,Paxton2013,Paxton2015,Paxton2017,paxton:19})
version 8118. We evolve a  15 $M_{\odot}$ star
with solar metallicity, starting at the
a zero-age main-sequence. The simulation continues 
until the central density approaches
$10^{10}$ g cm$^{-3}$. Then we extract the stellar profile when the star is carrying out core O-burning and map the stellar profile to our two-dimensional mesh. 

In the top left panel of Figure \ref{fig:evol} we plot a Hertzsrung-Russell diagram of the model's evolution, and in the top right panel of Figure \ref{fig:evol} we plot the time evolution of the nuclear, neutrino and surface luminosities of the stellar model. The surface luminosity  does not vary during the late-stage vigorous nuclear burning in the core, but the neutrino luminosity in general follows the nuclear luminosity. Note that the shell burning of C and Ne  triggers a brief jump in the nuclear luminosity, while O-burning provides a more steady energy production. In the bottom right panel of Figure \ref{fig:evol}, we plot the density and temperature  profile of the MESA profile. We model only the outer part beyond $10~R_{\odot}$ in the simulations, corresponding to $m(r)/M \sim 0.5$ at the base of the H-envelope where the wave heating is expected to take place.

\subsection{Hydrodynamics}

We use a two-dimensional hydrodynamics code which we previously
used extensively for SN simulations. The code is designed for multiple purposes
and has been used in simulations
of Type Ia SNe \citep{Leung2015a,Leung2018Chand,Leung2019subChand},
electron capture SNe \citep{Leung2017,Leung2019PASA,Leung2019ECSN,Zha2019}
and accretion-induced collapse \citep{Leung2019DMAIC,Zha2019DMAIC}.
The code utilizes the fifth-order shock capturing WENO 
\citep{Barth1999} and the five-step third-order 
non-strong-stability preserving Runge-Kutta scheme for time-discretization \citep{Wang2007} of the Euler equations. In this work, 
we consider the equatorial plane of the star
in polar coordinates ($\rho,\theta$). This allows
the use of a reflective boundary in the radial direction
and a periodic boundary in the $\theta$-direction. 
To perform the simulations in a computationally
feasible time range, we run the simulations in a single quadrant.


To close the equations, we use the Helmholtz equation of
state \citep{Timmes2000a}. This equation of states consists
of contributions from electron gas with arbitrary degeneracy,
ions in the form of an ideal gas, photons in the 
Planck distribution, and electron-positron pairs. 
We use a reflective boundary for 
the radial inner boundary, and a
free-flow boundary for the radial outer boundary.
When the energy deposition is strong, the fluid near the inner boundary can expand significantly, causing mass flow outwards.
In that case, we provide additional mass to the innermost grid 
cells such that the outflow mass is balanced by the additional
mass, as it would be in a real star by upwelling material. Then, we inject the corresponding gravitational energy to the system to maintain
energy conservation and adjust the mass cut value. This ensures that 
the total mass and energy, including both the hydrodynamical grids
and the mass-cut, are conserved as a whole.
 
\subsection{Energy Deposition}

Below we describe the changes to the code in order to accommodate
the wave heating physics. 
We consider a wave luminosity emerging from the core, $L_{\rm wave,0}$, which propagates into our computational domain and then dissipates into heat. We model only the heat dissipation and not the waves themselves.
In each time step,
we compute how much energy is deposited in each of
the computational mesh points by comparing the maximal
luminosity $L_{\rm max} = 2 \pi r^2 \rho c_s^3$
and the local $L_{\rm wave}$ which decreases with radius 
as wave energy is converted into heat. 
When $L_{\rm max} > L_{\rm wave}$,
no change in $L_{\rm wave}$ is made. Otherwise,
the new $L_{\rm wave}$ is computed by solving
(see e.g., \citealt{Landau1959,Ulmschneider1970,mihalas:84,Fuller2018} for derivations)
\begin{equation}
\label{Lwave}
\frac{d L_{\rm wave}}{dm} = \frac{\gamma + 1}{3 \pi} \omega c_s^2 \left( \frac{L_{\rm wave}}{L_{\rm max}}\right)^{3/2}.
\end{equation}
This formula is appropriate for waves that gradually steepen to form weak shocks (i.e., the post shock pressure only slightly exceeds the pre-shock pressure), which occurs when $L_{\rm wave}$ is slightly larger than $L_{\rm max}$. As discussed in \cite{Fuller2018}, the weak shock damping prevents the waves from steepening into strong shocks with $L_{\rm wave} \gg L_{\rm max}$, so the approximation of equation \ref{Lwave} is good throughout most of the wave energy deposition region.

The relative change $\Delta L_{\rm wave}/L_{\rm wave}$
can be very steep for a high value of $L_{\rm wave}$. We smooth by hand the energy deposition to a mass with a minimum of $m_{\rm dep,min} = 0.1~M_{\odot}$, which ensures the process of energy deposition
is resolved at the current resolution.
To do this, the energy along each ray with a given $\theta$ is smoothed over a mass
$m_{\rm dep,\theta} = 0.1 M_{\odot} / N_{\theta}$, where $N_{\theta}$ is the number
of grid points along the $\theta$-direction. 
In each step, we search along a constant $\theta$-direction
for the transition position ($r_0,\theta_0$) where $L_{\rm max} =
L_{\rm wave}$. Then we integrate outward 
to look for the outer radius $r_1$ so that the accumulated mass 
$\int_{r_0}^{r_1} {\rm vol}(r,\theta_0) \rho(r,\theta_0) = m_{\rm dep,\theta}$.
The local energy deposition in each grid cell is the mass of the cell divided by $m_{\rm dep,\theta}$.

\subsection{Initializing Hydrodynamics}

After evolving the stellar model, we take one snapshot during core O-burning, and map the density and pressure into our 2D simulation. The profile from the stellar evolution code obeys hydrostatic equilibrium. However, during mapping
from the Lagrangian MESA mesh to the Eulerian simulation mesh, slight departures from hydrostatic
equilibrium can be introduced. For the inner boundary, we chose a mass cut of mass $\sim 5.5~M_{\odot}$
with a radius $\sim 7.5 \times 10^{11}$ cm, near the base of the hydrogen envelope and below the initial wave heating region. We use a uniform 1500x30 grid with a radial grid size $1.5 \times 10^{11}$ cm and angular grid size of 3 $\deg$. 

In Appendix \ref{sec:static}, we further study how our model achieves hydrostatic equilibrium after it is mapped to our hydrodynamics model. While radial motion and changes to the density profile do occur, they are smaller than those caused by the wave heating which we examine below.
Additionally, convection is not present in the initial model, which may underestimate the energy transport at early times if  
convective flows can help channel the deposited energy away. 

The initial profile is spherically symmetric, so we add  
density perturbations in the form
$\delta \rho (r, \theta) = \alpha \rho \sin(12 \theta) \sin(b r /R)$. 
We choose $\alpha = 0.01$ in this work. 
In Appendix \ref{sec:Edepmass}
we demonstrate that the exact value of $\alpha$ does not change
our results. 
The density perturbation is added to mimic the inherent density fluctuations due to convection, and aims at seeding the growth of non-radial instabilities.
The value $b$ is chosen such that the total mass is conserved globally and along each radial direction. 
This prevents global motion of the star in any specific direction.

\begin{figure*}
\centering
\includegraphics*[width=18cm]{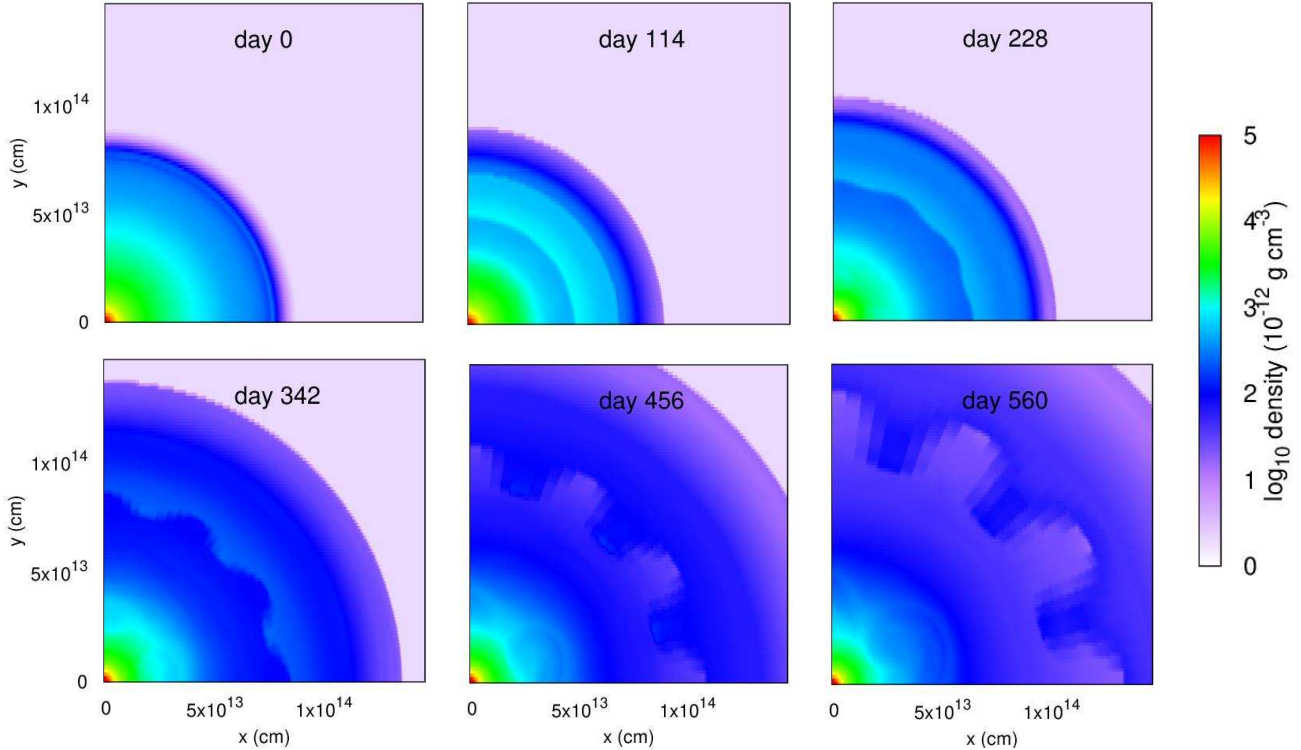}
\caption{
Density color plots of our benchmark model (Model 4 from Table \ref{table:models}, with $L_{\rm wave} = 3 \times 10^6 \, L_\odot$ for a duration of 120 days) at 0, 114, 228, 342, 456 and 570 days after the onset of energy deposition. The box size of all panels are fixed at $(1.5 \times 10^{14}, 1.5 \times 10^{14})$ cm for comparison. 
}
\label{fig:thermal_plot}
\end{figure*}

\section{Benchmark Model}
\label{sec:results}

Here we focus on this $15 \, M_\odot$ benchmark model, and in Section \ref{sec:discussions} we further discuss the dynamics based on models
of different progenitor masses.

\subsection{Hydrodynamical Response of Benchmark model}

When the energy deposition starts, the heated region is usually
within a narrow band of radius near the bottom of the simulation domain. As outlined in the previous
section, the deposition region is smoothed along 0.1 $M_{\odot}$
starting from the innermost grid point with energy deposition. For our benchmark model the heating rate is set to $3 \times 10^6 ~L_{\odot}$ and lasts for 120 days. 
In Figure \ref{fig:thermal_plot}, we show density color plots of the benchmark model 
at different time slices to show how the star expands and contorts due to energy deposition. 
The energy deposition creates a layered structure, due to weak shocks propagating through the envelope,
as seen in Day 114. Like the 1D models, the 2D models exhibit a ``bubble" of lower density within the envelope, but with a weaker density inversion. At day 114, the bubble surface is near $6 \times 10^{13}$ cm but expands outwards, and by day 228, the bubble surface is wavy due the density perturbations introduced in our initial conditions.
The heating ceases after day 120, but the star does not expand significantly until after day $\sim$200 when the heat-induced pressure wave reaches the stellar surface.
The core also develops non-radial structures due to non-radial instabilities and flows driven by the heating.

\begin{figure}
\centering
\includegraphics*[width=8cm,height=6cm]{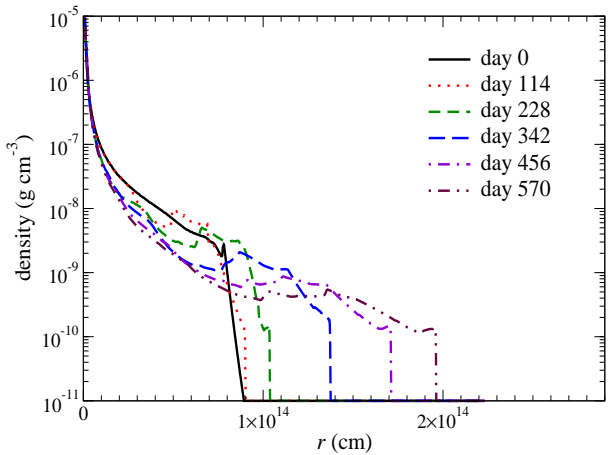}
\caption{
The angle-averaged density profiles of the benchmark model
at selected times after the start of energy deposition.
}
\label{fig:rho_profile_plot}
\end{figure}

We plot in Figure \ref{fig:rho_profile_plot} the angle-averaged density profiles
of the benchmark model at selected times
to further analyze the density and velocity evolution of the ejecta. The energy deposition takes place near the core at $\sim 10^{13}$ cm, driving a density inversion which propagates outward. The density difference can be a factor of $\sim$3 between the minimum value in the trough to the peak.
The angle-averaged density inversion gradually smooths out as the star expands, due to the growth of the Rayleigh-Taylor instability from the seed perturbations added to the model, which create the wavy patterns seen in Figure \ref{fig:thermal_plot}.

\begin{figure}
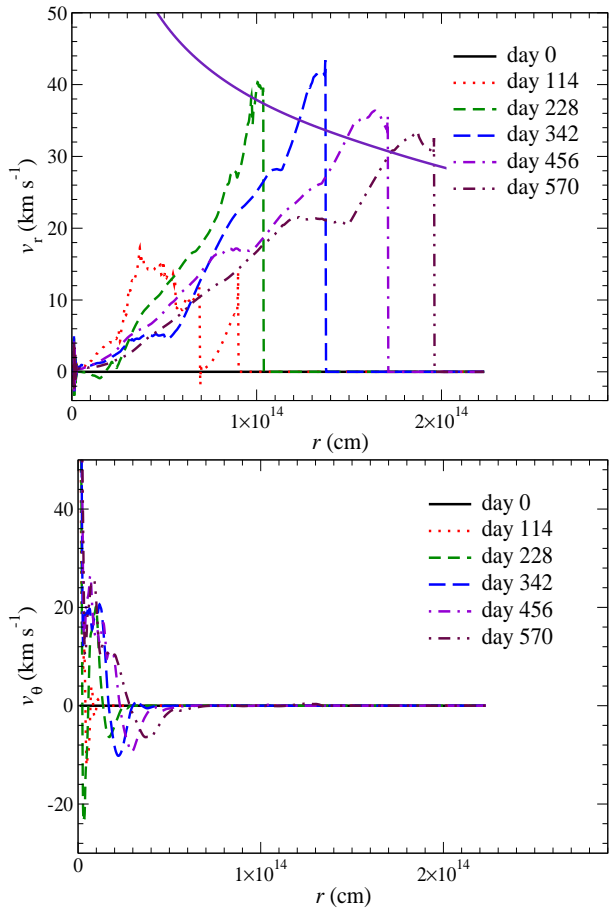

\centering
\includegraphics*[width=8cm,height=6cm]{velr_Model4_plot.eps}
\includegraphics*[width=8cm,height=6cm]{velz_Model4_plot.eps}
\caption{ {\bf Top:} The angle-averaged $v_r$ profiles of the benchmark model at selected times. The solid purple line corresponds to the escape velocity defined by  $v_{\rm esc} = (2 G M_{\rm int} / r)^{1/2}$, where $M_{\rm int}$ is the enclosed mass within a radius $r$. 
{\bf Bottom:} The angle-averaged $v_{\theta}$ profiles. 
}
\label{fig:vel_profile_plot}
\end{figure}

In the left panel of Figure \ref{fig:vel_profile_plot}, we plot 
the angle-averaged radial velocity
profile of the benchmark model.
The energy deposition occurring in the first 120 days drives expansion of the envelope, with internal velocities $\sim \! 15$ km s$^{-1}$. When this velocity/pressure wave reaches the surface where the density is small, it steepens into a shock to trigger a rapid expansion of the envelope. The velocity can be 
as high as $\sim 40$ km s$^{-1}$, near the escape speed of the star's surface. As the envelope expands, it slows down due to gravity, approaching $\sim 30$ km s$^{-1}$ at 570 days after the start of energy deposition. 

To further understand the asphericity of the ejecta, we also plot in the bottom panel of Figure \ref{fig:vel_profile_plot} the angle-averaged angular velocity snapshots. Most of the star has a smaller angular velocity compared to the radial velocity, but the angular velocity reaches $\sim 20$ km s$^{-1}$ within the innermost $2 \times 10^{13}$ cm.
The direction of the angular flow changes with both radius and time due to its turbulent nature. Beyond $\sim \! 5 \times 10^{13}$ cm, there is no observable motion in the angular velocity.

In this model, there is no direct mass ejection. The outer boundary 
at $2.3 \times 10^{14}$ cm is sufficiently large to 
contain all matter during the simulation.
Only the outer layers of the star achieve velocities close to the escape velocity. While pressure gradients may still provide extra
acceleration,
the sub-escape velocities of inner layers suggests the net mass ejection will be low. Besides, after the O-burning 
has started, it takes $\sim \! 1$ year before the final
collapse takes place, so there might not be sufficient time for any
matter to escape before the final collapse. 

\begin{figure}
\centering
\includegraphics*[width=8cm,height=6cm]{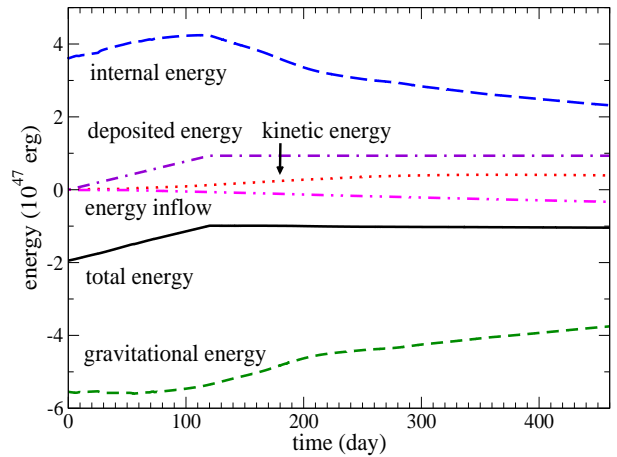}
\caption{
Energies as a function of time for the benchmark model. The energy components, including kinetic energy
(red dotted line), gravitational 
energy (green dashed line), internal energy (blue long-dashed line), deposited energy (purple dot-dash line), energy inflow (magenta dot-dot-dash line), and total energy (black solid line) are included.
}
\label{fig:energy_profile_plot}
\end{figure}

Finally, we examine the energy budget of the benchmark model. In Figure \ref{fig:energy_profile_plot}, we plot the different
components of energy as a function of time. The total energy grows linearly
in the first 120 days due to the deposited energy in the model.
Note also the small negative energy inflow due to the gravitational potential energy of mass added to the system as described in Section 2.
Examining the gravitational energy and internal energy trends, we see that the star does not significantly expand in the first 100 days, but then expands noticeably, lowering the gravitational energy. The early growth and subsequent decline of internal energy shows that the system first stores the deposited energy in thermal energy, and later converts this into kinetic energy, which is then converted into gravitational energy as the star expands outward. We also note that the internal energy eventually decreases below its initial value, reflecting the nearly adiabatic losses during envelope expansion.

\section{Dependence on Wave Heating Parameters}
\label{sec:massloss}

\begin{table*}
\begin{center}
\caption{
Parameters and energetics of the models studied
in this work. $L_{\rm wave}$ is the wave heating rate, Duration is the length of wave energy deposition, and $E_{\rm dep}$ is the total energy deposited. 
$E_{\rm ini}$ and $E_{\rm fin}$ are the initial and final total energy
of the envelopes in the simulations. }
\label{table:models}
\begin{tabular}{|c|c|c|c|c|c|c|c|}
\hline
Model & $L_{\rm wave}$ & Duration & $E_{\rm dep}$ & $E_{\rm ini}$ & $E_{\rm fin}$ & Unbound mass & remarks\\ \hline
Unit & $10^7 ~L_{\odot}$ & day & $10^{47}$ erg & 
$10^{47}$ erg & $10^{47}$ erg & $M_{\odot}$ & \\ \hline
1  & 0.1 & 120   & 0.398 & -1.948 & -1.586 & 0.04 & \\
2  & 0.1 & 360   & 1.195 & -1.948 & -1.033 & 1.3 & \\ \hline
3  & 0.3 & 40    & 0.398 & -1.948 & -1.585 & 0.01 &\\ 
4  & 0.3 & 120   & 1.195 & -1.948 & -1.055 & 0.26 &\\
5  & 0.3 & 360   & 3.585 & -1.948 &  0.686 & 3.9 & \\ \hline
6  & 1.0 & 40    & 1.195 & -1.948 & -1.153 & 0.18 &\\
7  & 1.0 & 120   & 3.585 & -1.948 &  0.351 & 5.8 &\\ \hline
4a & 0.3 & 120   & 1.195 & -1.948 & -1.049 & No  & seed $ = 0.1$ \\
4b & 0.3 & 120   & 1.195 & -1.948 & -0.975 & No  & seed $ = 0.001$ \\ 
4c & 0.3 & 120   & 1.195 & -1.948 & -1.053 & No  & $M_{\rm dep} = 0.05 ~M_{\odot}$\\ 
4d & 0.3 & 120   & 1.195 & -1.948 & -1.046 & No  & $M_{\rm dep} = 0.2 ~M_{\odot}$ \\ 
4s & 0.3 & 120   & 1.195 & -1.948 & -0.685 & No  & No initial seed\\ 
\hline
\end{tabular}
\end{center}
\end{table*}

In the previous section, we examined how the envelope responds to wave energy deposition for a benchmark model. Here, we extend our study to examine how these properties depend on the energy deposition rate and duration. Table \ref{table:models} lists parameters of the models we have run, where the benchmark model is Model 4. We also remark that, another way to compare the model
is by the total energy deposited. In this sense, Models 1 and 3 form
the ``low-energy model" (LEM); 
Models 2, 4 and 6 form the ``medium-energy models" (MEM); 
and Models 5 and 7 form the ``high-energy models" (HEM).

\subsection{Dependence on Total Wave Heat}

We first study how models with the same 
total deposited energy behave, from which we can 
extract some general observations about the effects of wave energy deposition.

\subsubsection{Hydrodynamics of Low-Energy Models}

\begin{figure}
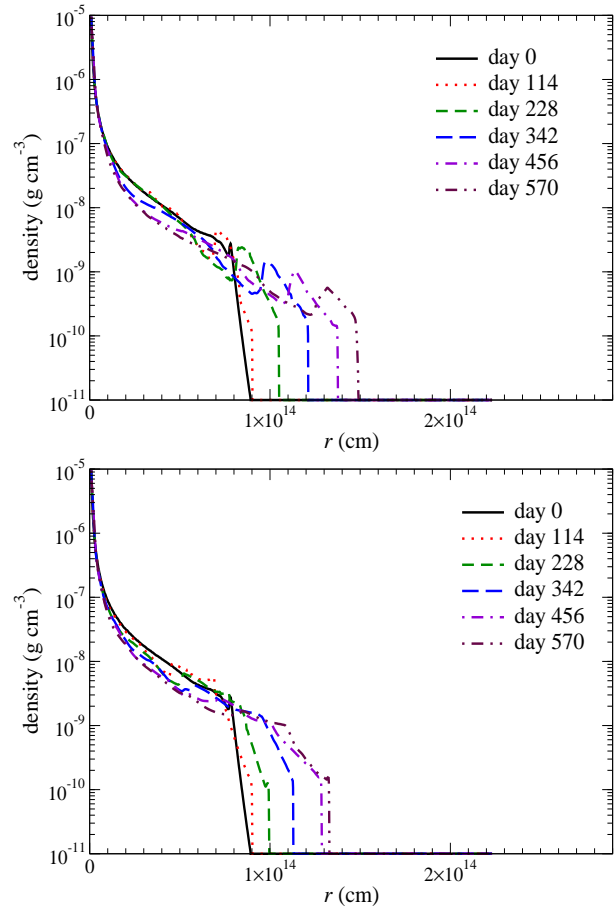

\centering
\includegraphics*[width=8cm,height=6cm]{rho_Model1_plot.eps}
\includegraphics*[width=8cm,height=6cm]{rho_Model3_plot.eps}
\caption{
The angle-averaged density profiles of Low-Energy Models 
($E_{\rm dep} \sim 4 \times 10^{46}$ erg) with
Model 1 (top) and Model 3 (bottom). The corresponding heating rates are $10^6 L_{\odot}$ for 120 days in Model 1 and $3 \times 10^6 L_{\odot}$ for 40 days in Model 3.
}
\label{fig:rho_isoE_time_plot1}
\end{figure}

In Figure \ref{fig:rho_isoE_time_plot1}, we plot in the upper (lower)
panel the density profiles of Model 1 (3) taken from the LEM.
In Model 1, which has a lower $L_{\rm wave}$, the star shows 
a regulated expansion. A density inversion (like that shown in
the benchmark model) develops at $8 \times 10^{13}$ cm at day 114 and slowly propagates to $2 \times 10^{14}$ cm by day 570.
The density contrast across the inversion nearly stays fixed at roughly a few times the minimum density, while the sharp density gradient at the surface remains.
In Model 3, the expansion proceeds steadily without forming a
density inversion. A strong shock does not form in either case, leaving a sharp density gradient at the surface, and all material in the star remains gravitationally bound.

\begin{figure}
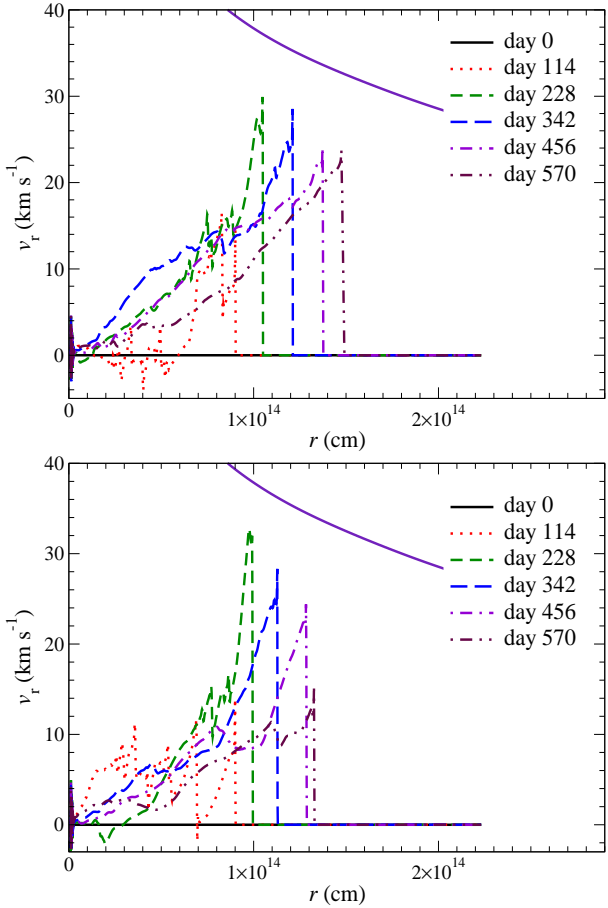

\centering
\includegraphics*[width=8cm,height=6cm]{velr_Model1_plot.eps}
\includegraphics*[width=8cm,height=6cm]{velr_Model3_plot.eps}
\caption{
{\bf Top:} The angle-averaged radial velocity profiles of Low-Energy Models ($E_{\rm dep} \sim 4 \times 10^{46}$ erg) with Model 1 (top) and Model 3 (bottom).
}
\label{fig:vel_isoE_time_plot1}
\end{figure}

Figure \ref{fig:vel_isoE_time_plot1} shows the radial velocity profiles of the LEM. In Model 1, the initial energy deposition triggers a pulse-like velocity profile at day 114 with a peak velocity of $\sim 30$ km s$^{-1}$ at day 228. The whole star is gravitationally bound, including the shock-heated surface.
Model 3 shows an erratic velocity profile characterized by shocks at early times, which can be seen at 3, 5, 7 and $9 \times 10^{13}$ cm at day 114. After the shocks reach the surface, the star exhibits nearly ``homologous" expansion in the sense that the expansion velocity is roughly proportional to radius. Again the velocity is too low to escape the star directly. The surface velocity is only 15 km s$^{-1}$ at the end of the simulation.
 
These two models suggest that the actual changes in the 
density and velocity profiles are almost negligible
in the low energy regime. The heat triggers moderate expansion but no major restructuring of the density profile.

\subsubsection{Hydrodynamics of Medium-Energy Models}

\begin{figure}
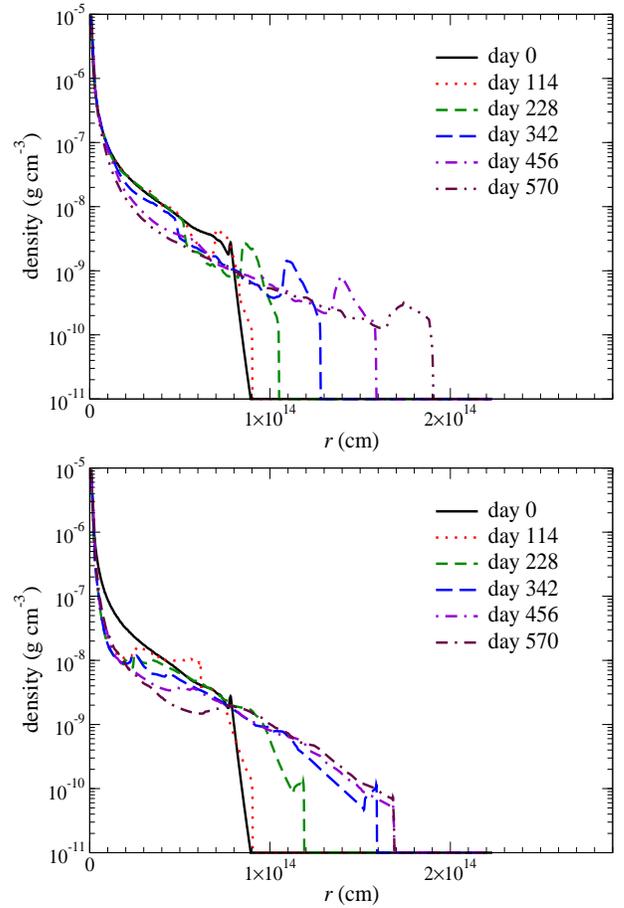

\centering
\includegraphics*[width=8cm,height=6cm]{rho_Model2_plot.eps}
\includegraphics*[width=8cm,height=6cm]{rho_Model6_plot.eps}
\caption{
{\bf Top:} Same as Figure \ref{fig:rho_isoE_time_plot1} but for the Medium-Energy Models ($E_{\rm dep} \sim 1 \times 10^{47}$ erg) with Model 2 (top) and Model 6 (bottom). The two models and the benchmark model share the same amount of deposited energy, but with a low heating rate of $10^6~ L_{\odot}$ for 360 days in Model 2 and a high heating rate of $10^7~ L_{\odot}$ for 40 days in Model 6.
}
\label{fig:rho_isoE_time_plot2}
\end{figure}

In Figure \ref{fig:rho_isoE_time_plot2} we plot the density profiles of two models with the same energy input as the benchmark model, with $E_{\rm dep} = 1.195 \times 10^{47}$ erg, but at a low heating rate of $10^6 ~L_{\odot}$ for 360 days (Model 2) and 
at a high heating rate of $10^7~ L_{\odot}$ for 40 days (Model 6). The lower energy deposition in Model 2 creates a 
density inversion which mildly accelerates with time as the density contrast approaches the surface. The density difference can be a few times higher than the minimum density at the trough. 
The star smoothly expands to $\sim 1.9 \times 10^{14}$ cm.  
In Model 6, there is no significant density inversion developed
in the whole simulation, except a small one near the surface. The stellar density profile is mostly monotonically decreasing and expanding, signifying a smooth expansion. The outer layers stagnate at the end of the simulation, suggesting that the star's expansion slows down, with all of its matter remaining bound.

\begin{figure}
\centering
\includegraphics*[width=8cm,height=6cm]{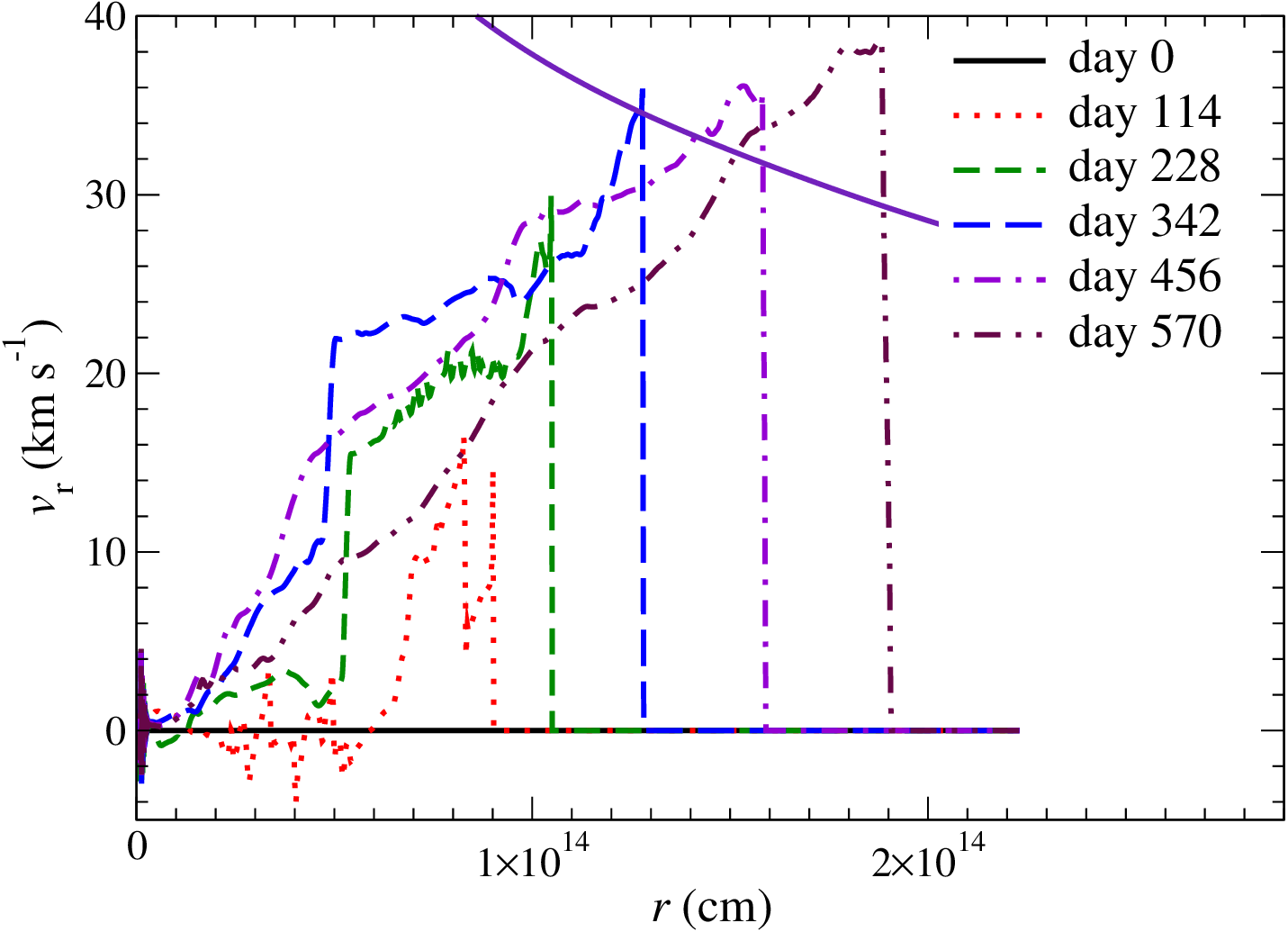}
\includegraphics*[width=8cm,height=6cm]{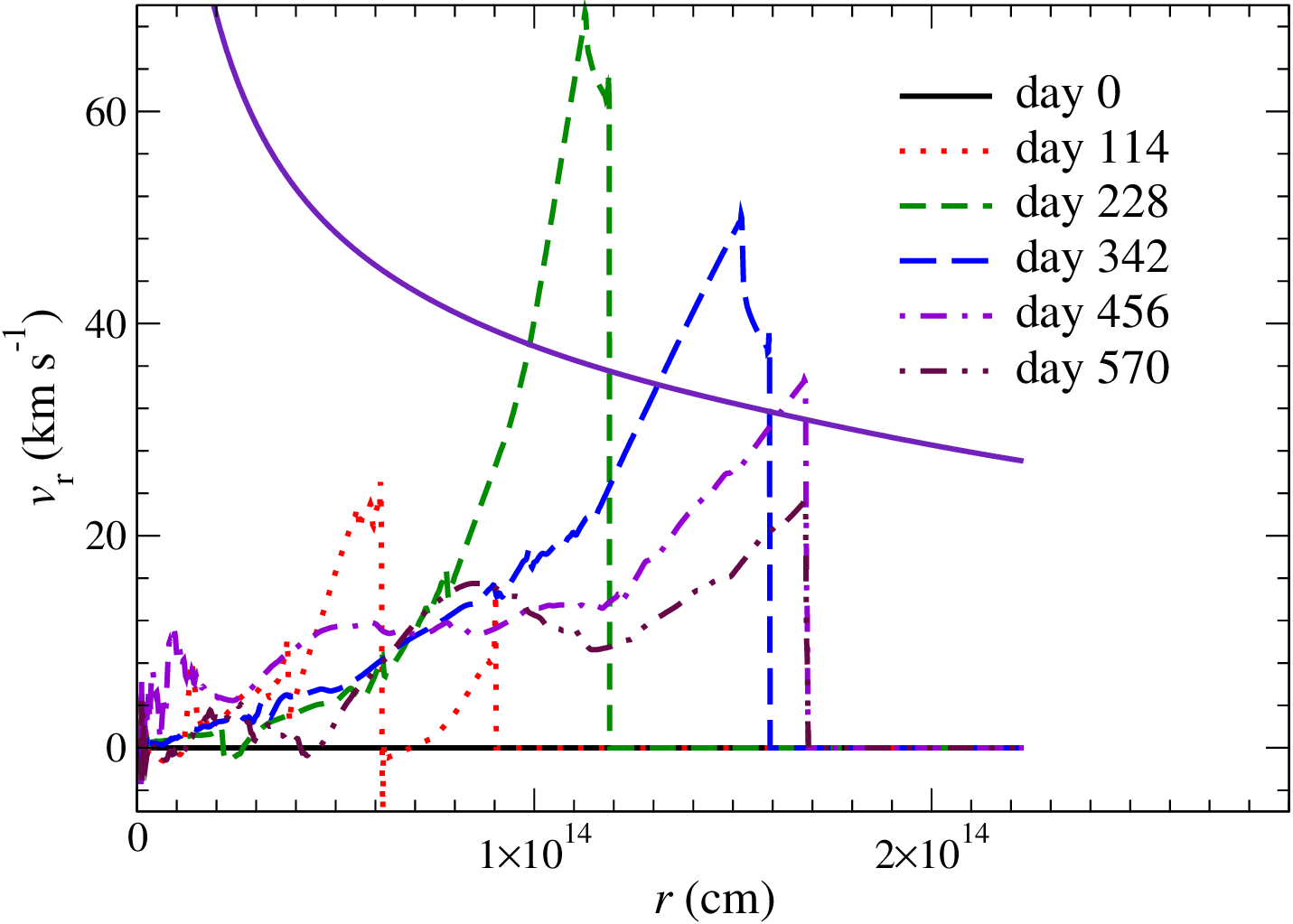}
\caption{
{\bf Top:} Same as Figure \ref{fig:vel_isoE_time_plot1} but for the Medium-Energy Models with Model 2 (top) and Model 6 (bottom).
}
\label{fig:vel_isoE_time_plot2}
\end{figure}

In Figure \ref{fig:vel_isoE_time_plot2}, we plot the velocity profiles of the MEM. Similar to the density profiles, the velocity
profiles also show interesting differences between the two models. Model 2 with a low heating rate does not show deceleration, and the star reaches nearly homologous expansion at the end of simulation. Model 6 shows almost the opposite behavior. The energy injection creates a two-peak shock structure at day 114,
with the inner shock traveling faster than the outer. The shocks merge into a stronger shock at day 228 that reaches above 60 km s$^{-1}$, almost 50 \% higher than Model 2, despite the same total energy deposited. However, the expansion is not sustained by further heating, and the surface velocity drops by half and reaches $\sim 30$ km s$^{-1}$ at the end of the simulation.


\subsubsection{Hydrodynamics of High-Energy Models}

\begin{figure}
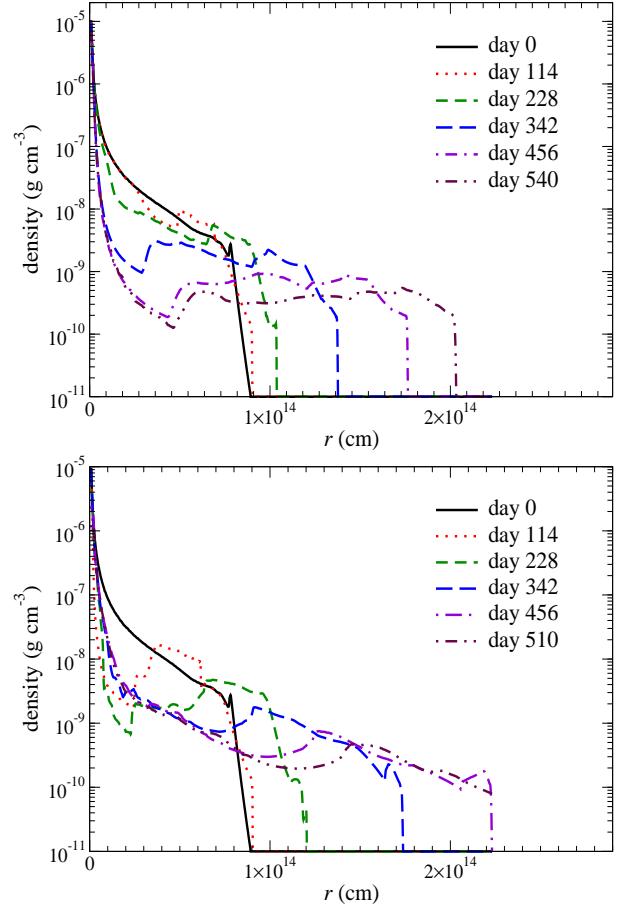

\centering
\includegraphics*[width=8cm,height=6cm]{rho_Model5_plot.eps}
\includegraphics*[width=8cm,height=6cm]{rho_Model7_plot.eps}
\caption{
{\bf Top:} Same as Figure \ref{fig:rho_isoE_time_plot1} but for High-Energy Models ($E_{\rm dep} \sim 4 \times 10^{47}$ erg) with Model 5 (top) and Model 7 (bottom). The two models have heating rates of $L_{\rm wave}=3 \times 10^6 \, L_\odot$ (Model 5) and 
$9 \times 10^6 \, L_\odot$ (Model 7).
}
\label{fig:rho_isoE_time_plot3}
\end{figure}

We next examine the HEM, Models 5 and 7, which have a higher deposited energy of $3.585 \times 10^{47}$ erg. Figure \ref{fig:rho_isoE_time_plot3} shows that the density profiles are drastically altered by the large energy deposition. Both models develop large density inversions as material in the inner envelope is lifted outwards. In Model 5, which has a lower $L_{\rm wave}$, the density profile is flattened, with a density peak near $6 \times 10^{13}$ cm at day 114 that propagates to nearly $2 \times 10^{14}$ cm by day 570. 
In Model 7, the higher heating rate drives a stronger shock that expels material to larger radii.
At day 456, the outer layers of the star reach the simulation outer boundary at $\sim 2.3 \times 10^{14}$ cm. 

\begin{figure}
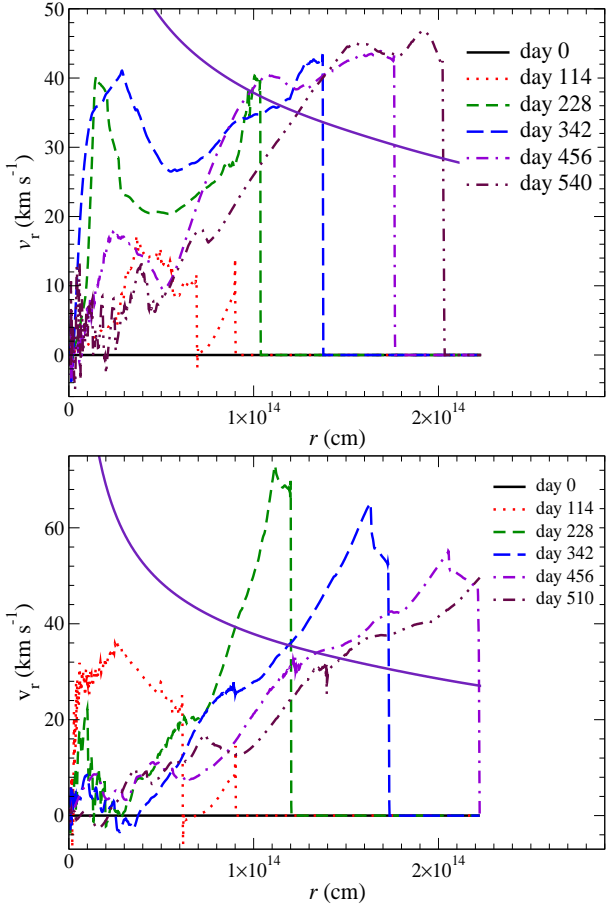

\centering
\includegraphics*[width=8cm,height=6cm]{velr_Model5_plot.eps}
\includegraphics*[width=8cm,height=6cm]{velr_Model7_plot.eps}
\caption{
{\bf Top:} Same as Figure \ref{fig:rho_isoE_time_plot3} but for High-Energy Models with Model 5 (top) and Model 7 (bottom). 
}
\label{fig:vel_isoE_time_plot3}
\end{figure}

In Figure \ref{fig:vel_isoE_time_plot3}, we plot the 
velocity profiles of Models 5 and 7 from the HEM series.
Model 5 shows that the star can still expand rapidly with a moderate $L_{\rm dep}$ but with a long duration. The velocity profile features a strong two peak structure at intermediate times but approaches homologous expansion at late times. The outer layers asymptotically approach a velocity $\sim 40$ km s$^{-1}$, larger than the escape speed.
The inner layers slows down with time, developing a complex turbulent structure in the core.

Dynamical features accompany the expansion in the velocity profiles of Model 7.
When the strong shock breaks out from the surface of the star around day 228, the velocity peaks at 70 km s$^{-1}$, compared to the local escape velocity of $\sim 40$ km s$^{-}$. Afterwards, the ejecta slows down but can still escape. Turbulent motion persists in the inner part of the star. The outer layers of the ejecta reach the outer boundary of the simulation as shown in the profile at day 456. At the end of the run, the star approaches homologous expansion.

\subsection{Dependence on Wave Heating Rate and Duration}

In this Section, we discuss general differences between models
with the same wave heating rate but different duration, 
or vice versa. The duration is determined by the lifetime
of the core burning. The duration of the advanced 
burning, such as O-burning, generally decreases with stellar mass.
The exact heating rate is also unclear as there is significant uncertainty in the wave luminosity generated by the convective core, but it is expected to increase for more massive stellar cores. Here, we explore the consequences of
the variation of these values on the mass ejection process. 

We examine how $L_{\rm wave}$ affects the evolution of the velocity profile by analyzing the three Models with a heating duration of 120 days: Model 1 (LEM), Model 4 (MEM), and Model 7 (HEM). The energy deposition rate significantly affects how the star expands and how the star ejects mass. A lower $L_{\rm wave}$,
makes the star expand but with no mass ejection. Higher values of $L_{\rm wave}$ can drive stronger shocks that lead to mass ejection from the surface. Only Model 7 with $L_{\rm wave} = 10^7 ~L_{\odot}$ exhibits significant mass ejection due to wave heating. The larger energy deposition of the HEM not only makes the star expand faster, but it also creates a prolonged trough in the density profile inside the star. In contrast, the LEM exhibits very little internal motion or mass ejection in its velocity profile.

We identify significant differences in the density and velocity structures between models with different heating rates. Models with low but sustained heating rates exhibit more steady expansion, with a small density inversion forming as the outer layers are gradually lifted outwards by the expanding inner layers. Due to the low heating rate, no strong shock wave is excited, and matter is not shock-accelerated near the surface. We refer to this as a ``bubble" phenomenon. In contrast, a larger heating rate excites a stronger shock that accelerates the surface layers outward, creating a low-density tail of material above the previous location of the photosphere, with a much smaller density inversion interior to this tail. We refer to this as a ``bomb" phenomenon. Thus, low heating rates create a ``bubble" phenomenon, while high heating rates (i.e., a heating time scale shorter than the local dynamical time scale) create a ``bomb" phenomenon.

\subsection{Global Comparison of Mass Loss}

\begin{figure}
\centering
\includegraphics*[width=8cm,height=6cm]{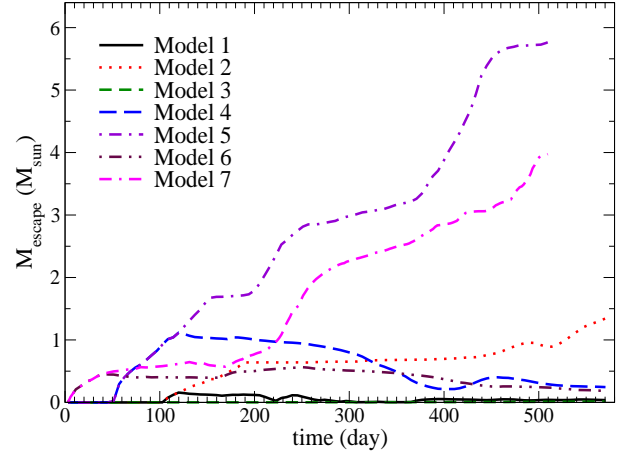}
\caption{The time evolution of the ejecta mass of models presented
in this work. The ejecta mass is defined by the total mass of matter where the sum of the local kinetic, internal and potential energy is positive.}
\label{fig:Mej_model_plot}
\end{figure}

In Figure \ref{fig:Mej_model_plot}, we plot the total ejecta mass 
against time for the models presented. The ejecta mass is defined
by the integrated mass elements when the sum of the local kinetic energy, internal energy and the potential energy is positive. This quantity is therefore time-dependent as demonstrated in previous sections.
The 7 models can be characterized into three groups based on the total energy deposition: LEM models exhibit little to no mass loss, MEM models exhibit moderate amounts of mass loss, and HEM models exhibit large amounts of mass loss.
For LEM (Models 1 and 3), effectively there is no 
ejected mass as the escape mass is as low as $\sim 10^{-2} ~M_{\odot}$.
The MEM (Models 2, 4, and 6) shows the ejected mass is roughly an order of magnitude
higher at 0.5--1.5 $M_{\odot}$. At last in HEM (Models 5 and 7),
the ejected mass reaches asymptotic values of 4--6 $M_{\odot}$, where the whole envelope mass is also $\sim 6 ~M_{\odot}$.

The heating rate also affects the evolution of the ejected mass. Unsurprisingly, models with the largest heating rates (Models 6 and 7) show the fastest initial increase in unbound mass, while models with the lower heating rates require a longer time for any mass to accumulate enough energy to escape. Between models with the same energy deposition, lower heating rates with longer durations generally produce larger ejecta masses.

\subsection{Global Comparison of Deposition Radius}

\begin{figure}
\centering
\includegraphics*[width=8cm,height=6cm]{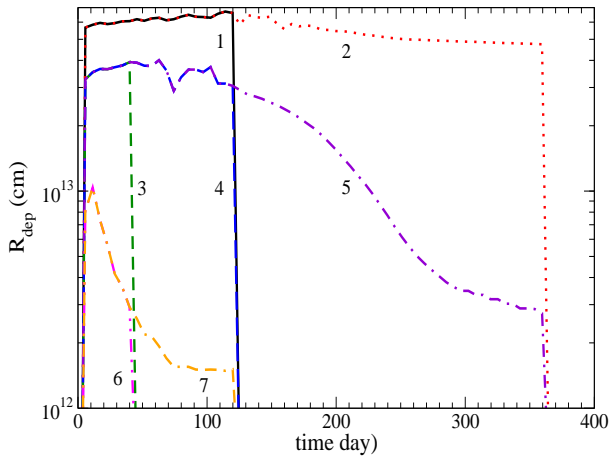}
\caption{The time evolution of the energy deposition radius for the models presented in this work. For each model, the deposition radius falls to zero when we turn off wave heat. In models with high heating rates (Models 5 and 7), the deposition radius moves inward as the density near the base of the simulation decreases due to wave heating.}
\label{fig:Rdep_model_plot}
\end{figure}

Here we compare the position where the acoustic waves 
deposit energy. In Figure \ref{fig:Rdep_model_plot}, we plot the position of the 
energy deposition against time for the models presented.
The sharp drop in all curves occurs when heating abruptly shuts off. We see that models with a lower $L_{\rm dep}$ 
have their energy deposition at larger radii,
which differ from $\sim 5 \times 10^{13}$ cm in Models 1 and 2,
down to $2 \times 10^{12}$ cm for Model 7. As the 
acoustic waves deposit their energy only when the maximum wave
energy flux is exceeded (equation \ref{Lwave}), a lower energy deposition rate requires a lower density and sound speed to match this criteria, so the heating occurs at larger radii. The roughly constant heating radius in Models 1--5 before day 100
suggests the wave energy does not strongly alter the stellar structure in the wave heating region during this time.
The trends in the heating radius are correlated with $L_{\rm dep}$, because
the wave heat causes the inner envelope to expand and decrease in density, causing the heating radius to move inward. This happens almost immediately for the largest heating rates. As seen in Models 5 and 7, the recession of the deposition sphere can reduce its size by almost an order of magnitude.

\subsection{Global Comparison of Energy}

\begin{figure}
\centering
\includegraphics*[width=8cm,height=6cm]{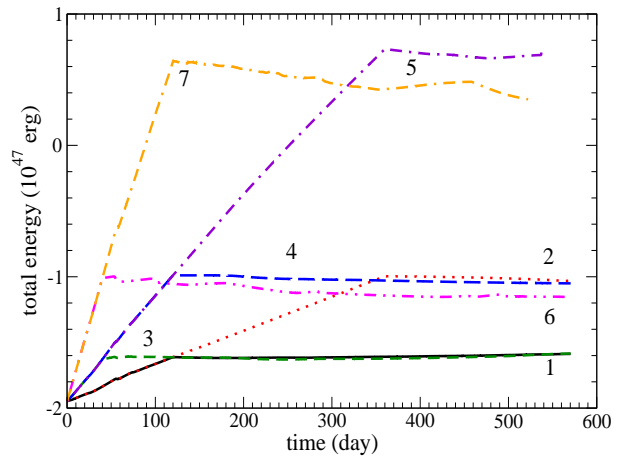}
\caption{The time evolution of the total energy
for the models presented in this work.}
\label{fig:energy_model_plot}
\end{figure}

At last, we compare the total energy as a function of 
time among all models, plotted in Figure \ref{fig:energy_model_plot}. Unsurprisingly, the final energy is mostly determined by the total acoustic wave energy deposition. Note that only Models 5 and 7 finish with positive total energy, implying large amounts of unbound material, as verified by the results in Figure \ref{fig:Mej_model_plot}. However, as the star expands, extra contributions of energy emerge due to the energy content of inflowing/outflowing matter across the simulation boundaries, so the final energy in the simulation is not merely the sum of the initial energy  and the heating energy.

This is not just an artifact of the simulations, as mass in a real star will generally flow from below the wave heating region to above it, changing the net energy content of the envelope. This effect is negligible for models with lower $L_{\rm dep}$, due to a smaller contribution from material flowing across the boundaries. Models with larger heating rates drive material away from the inner boundary, causing more matter to flow into the simulation domain from the inner boundary, which advects negative gravitational energy into the simulation domain. 

In model 5, there is a mild increase of energy between day 400-450.  This is the period when the rapid expansion creates an energy discontinuity near the isothermal outer layer, which is set to be at the minimum temperature allowed in the simulation ($\sim 3000$ K). When the discontinuity passes across this region, fluid elements can be cooled below the minimum temperature. In this code we choose to stick with thermodynamical consistency instead of energy consistency, so the missing energy between the actual temperature and the temperature floor is added into the system. Since this only alters the total energy budget by about $1 \%$, we do not believe it qualitatively affects our conclusions.



\section{Discussion}
\label{sec:discussions}

\subsection{Pre-supernova Mass Loss in Stars of Other Masses}

\begin{figure*}
\centering
\includegraphics*[width=18cm]{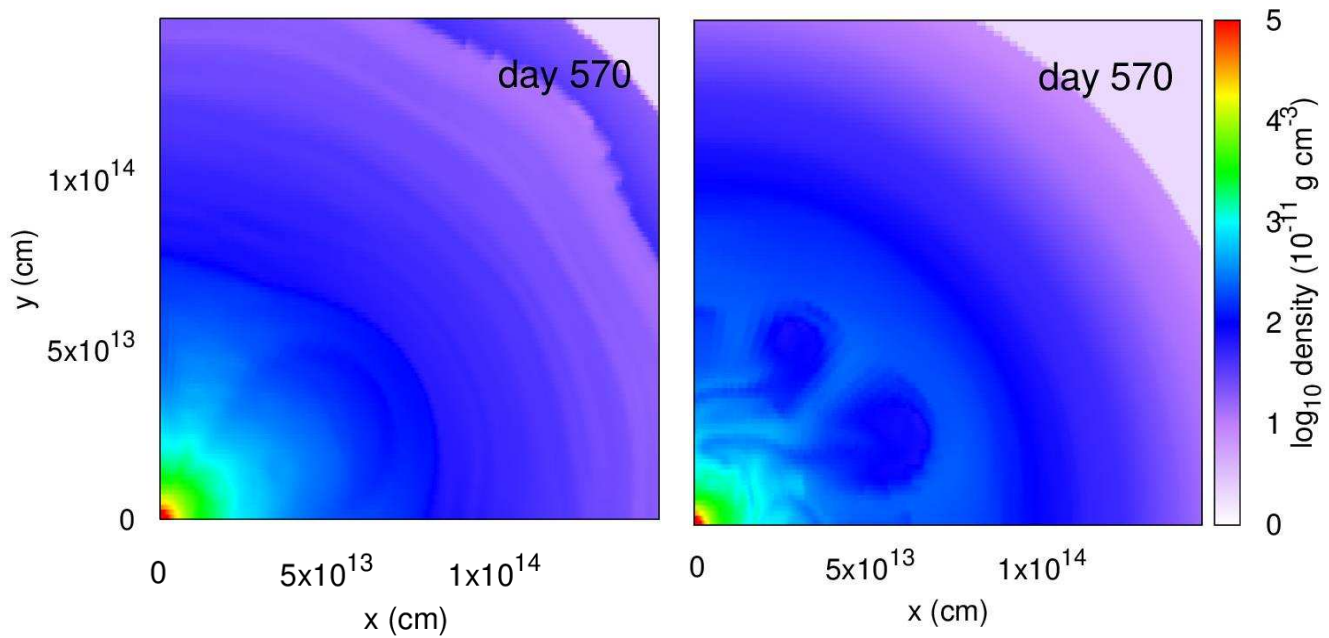}
\caption{
{\bf Left:} Density distributions of Model 2 (left) and Model 6 (right)
at the end of the simulation. The two models share the 
same total energy deposition of $\sim 1.2 \times 10^{47}$ erg,
but with $L_{\rm wave} = 10^6 ~L_{\odot}$ for 12 months in Model 2
and $L_{\rm wave} = 10^7 ~L_{\odot}$ for 4/3 months in Model 6.}
\label{fig:asphericity}
\end{figure*}

In general, the outburst process depends strongly on the wave excitation in the core, in addition to the properties of the envelope, and both core and envelope should be evolved together for robust estimates. 
Fortunately, most red supergiant envelopes have very similar, nearly poltropic structures. The envelope masses are also similar, $\sim \! 7 \, M_\odot$, since higher mass stars have more of their mass in their cores, and lose more envelope mass by winds. In contrast, stars of different mass have very different wave heating rates and heating durations \citep{shiode:14}. Hence, in our parameter study, lower values of $L_{\rm wave}$ for longer durations are more representative of low-mass stars, while higher values of $L_{\rm wave}$ for shorter durations are more representative of high-mass stars. Different burning phases behave similarly: carbon shell-burning can be much longer ($\sim \! 10^2$ years) but the wave luminosity is correspondingly smaller. Silicon burning generates much larger wave fluxes but only lasts days, so the envelope does not have time to respond \cite{Fuller2015}.

It is possible that other processes occur in some stars, such as convective shell mergers. These would spur intense nuclear burning and vigorous convection \citep{Meakin2006,Collins2018,Yadav2019}, driving very large wave fluxes into the envelope. Another scenario is that the core dynamo amplifies the magnetic field in the convective shells near the core, which could also generate waves and lead to energy deposition near the surface \citep{Soker2017}.
While we do not investigate such processes here, an uncommon subset of stars could generate wave fluxes much greater than our fiducial estimates.

\subsection{Appearance of Wave-driven Outbursts}

Our simulations do not include radiative transfer and cannot predict the detailed appearance of wave-driven pre-SN outbursts. From the 1D models of \cite{Fuller2017}, the photometric signal of a pre-SN outburst in a red supergiant is likely dominated by shocks propagating up to the photosphere after short but intense periods of wave heating, such as the core neon burning phase. Our 2D simulations with large heating rates also launch these shocks, which are not strongly affected by non-radial instabilities. Hence, we expect that the observable features of pre-SN outbursts are captured reasonably well by 1D models. 
When the outburst takes place in a binary system, the accretion of the ejected matter by the companion star might also produce a bright event \citep{Mcley2014,Danieli2019}.

\subsection{Envelope Asphericity}

\begin{figure*}
\centering
\includegraphics*[width=18cm]{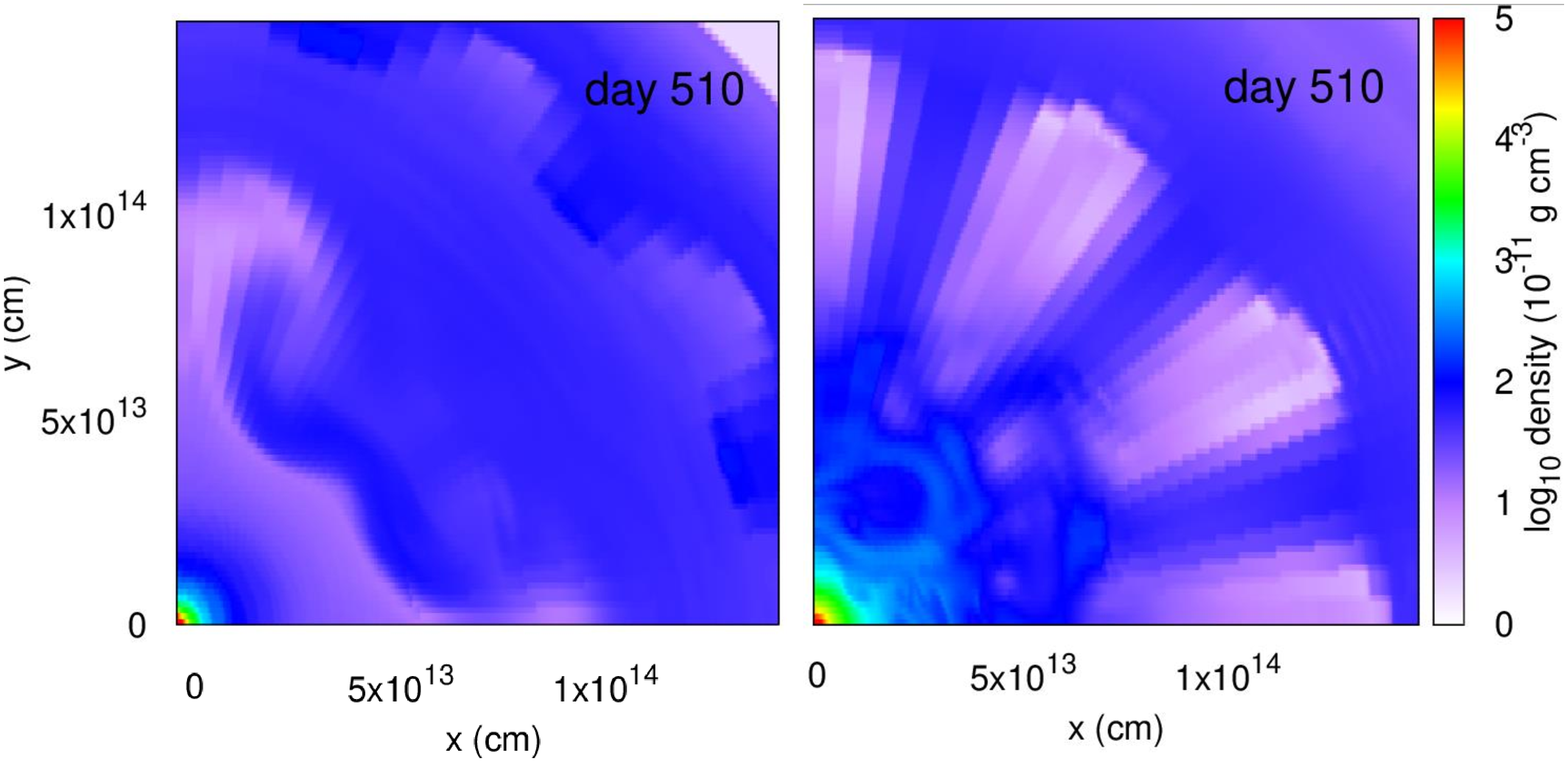}
\caption{
{\bf Left:} Density color plot of Models 5 at the end of the simulation.
{\bf Right:} Same as top panel but for Model 7.
The two models share the 
same total energy deposition of $\sim 3.7 \times 10^{47}$ erg,
but with $L_{\rm wave} = 3 \times 10^6 ~L_{\odot}$ for 12 months in Model 5
and $L_{\rm wave} = 10^7 ~L_{\odot}$ for 4 months in Model 7.}
\label{fig:asphericity2}
\end{figure*}

In this work, we have extensively computed two dimensional models
showing how the wave heating generates asphericity and 
drives mass loss. In Figure \ref{fig:asphericity},
we further demonstrate how the asphericity develops at late times. We examine Models 2 and 6,
finding that both models are spherical to a good approximation. However, notable turbulence is driven within the middle of the envelope, which is especially evident at at radii around $5 \times 10^{13}$ cm. These structures are much more pronounced in Model 6, likely because the larger heating rate (but shorter duration) fuels more expansion of the inner envelope, generating stronger Rayleigh-Taylor instability.
Furthermore, the density profile in Model 2 shows
observable ``wrinkles", though it is not yet clear if these features are real or numerical artifacts. 
For LEM and MEM, the asphericity is small, and the wave heating primarily contributes to a quasi-spherical expansion
of the stellar envelope.

For comparison, we also plot in Figure \ref{fig:asphericity2}
the density color plots of two higher energy models, Models 5 and 7.
The most striking features of these models is the large-scale structure with three wavelengths per quadrant at radii around $10^{14}$ cm. This is clearly a consequence of the initial configuration of the simulations, where a sinusoidal perturbation with three periods in the radial and angular directions (and an amplitude $1 \%$) was embedded in the density profile. However, by the end of the simulation, this $1\%$ perturbation has grown to a large amplitude, producing as much as a factor of 10 variation in the horizontal density fluctuations. We attribute this large amplification to rapidly growing Rayleigh-Taylor instability. Interestingly, the radial location where the instability is most prominent depends on the wave heating rate, but it is clear that higher total heat deposition allows for more growth of density perturbations. Of course, a real star would have initial density fluctuations at a variety of scales due to turbulent convection in the envelope, but these models indicate those perturbations can be magnified by wave heating effects, potentially producing significant asphericity in the exploding star. 

To quantify the level of asphericity in the models, 
we compute define the mass-weighted asphericity via the Fourier modes
\begin{equation}
S_n = \frac{1}{M} \int_{R_{\rm cut}}^{R} \int_{0}^{\pi/2} 8 r^2 dr d\theta~\rho \sin (n \theta) \, .
\end{equation}
Here, the integer $n$ is the Fourier harmonic and $M$ is the total mass of the star. We also compute a similar expression with $C_n$ using $\cos (n \theta)$ as the Fourier component, and we thus define the asphericity of $n$-th order, $A_n = \sqrt{C_n^2 + S_n^2}$.

\begin{figure}
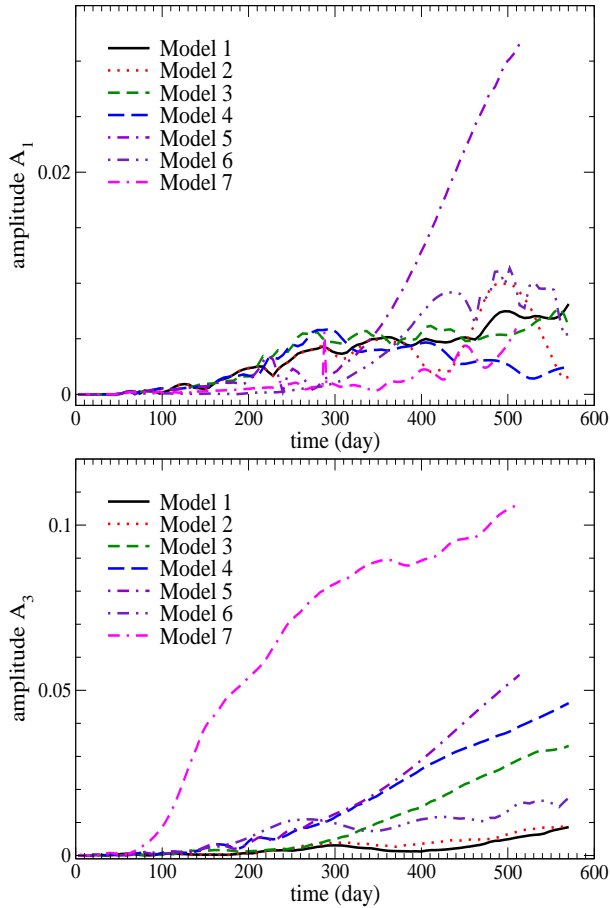

\centering
\includegraphics*[width=8cm,height=6cm]{amplitude_A1_time_plot.eps}
\includegraphics*[width=8cm,height=6cm]{amplitude_A3_time_plot.eps}
\caption{
The time evolution of the density asphericity
for all 7 models presented in the main text. The asphericity
is defined by the $A_n = \sqrt{C_n^2 + S_n^2}$ where 
$C_n$ and $S_n$ are the cosine and sine components of the angular Fourier
modes of the density profiles. 
The top panel shows the elongation $A_1$ while the bottom panel shows the $A_3$ component.
}
\label{fig:xi_time_plot}
\end{figure}

In Figure \ref{fig:xi_time_plot} we plot the time evolution of the asphericity modes $A_1$ and $A_3$ of all 7 models
presented in previous sections. The $n = 3$ component is often dominant as discussed above. Comparing the two plots, $A_1$ is indeed smaller than $A_3$ by a factor of a few.
The LEM show a mild increase in asphericity, but their low values show the stars remain nearly spherical. The MEM show a larger asphericity that grows nearly linearly in time. The
HEM have the most significant asphericity, with the $A_3$ component of Model 7 reaching 10 \%. Despite its lower total energy deposition, Model 5 exhibits the largest value of $A_1$ and hence has the most elongated structure, though the elongation only reaches $\approx 3 \%$. The angular dependence of the envelope density may give rise to interesting effects for the subsequent SN and its appearance.

In Appendix \ref{appendix:seed} we further compare models with different initial aspherical seeds to see how this affects the star's final asphericity, finding the results are not strongly dependent on this choice.

\subsection{Limitations of This Work}


First, our simulations use a moderate resolution which is almost constant
throughout the star with $\Delta x = 1.5 \times 10^{10}$ cm.
This $\Delta x$ is rather large compared to the mass cut
which is $\sim 7.5 \times 10^{10}$ cm. Thus, the inner features
of the star might not be fully captured and may be under-resolved.
This value of $\Delta x$ can provide us the necessary resolution for 
the outer part to keep track of the smaller features due to 
hydrodynamical instabilities. Most such features should be produced in the outer envelope as that is the place where the density inversion is present.

Second, our simulation does not resolve the acoustic waves emitted from the core, instead only including a heating term to account for their energy deposition. Studying how these waves deposit heat and momentum is numerically challenging because it takes place on a short length and time scale, i.e., the wavelength and wave period. The huge number of acoustic waves leaked into the envelope provides the possibility of modeling this process statistically, namely by the recognition of $L_{\rm wave}$ as a time-averaged quantity. Interestingly, our results imply the net effect is for acoustic waves to damp via weak shocks, converting their energy into pressure pulses with larger length and time scales, which then continue to propagate out until they shock again. It would be an interesting extension to resolve both the original acoustic waves and the secondary pressure pulses to see if the results differ from the treatment adopted here.

Third we have used the Helmholtz equation of state for our computation.
But for modeling the envelope, we are approaching a density
close to $10^{-11}$ g cm$^{-3}$, which is the density floor provided
by the equation of state table. As discussed above, the temperature would also decrease below $10^4$ K if we did not impose a temperature floor. 
This limits us to explore further how the ejecta behaves beyond day 570. In some of our more energetic models, vacuum zones develop inside the star, with a few grid cells along the same $r$ reaching the density floor. These are not evident in the figures because the density profiles shown are angle-averaged. The large associated pressure differences begins to trigger unphysical flow. 
Further extrapolation to lower densities and temperatures will require an updated
equation of state. Additionally, at these low densities, the matter might not be tightly coupled to the radiation field, so non-diffusive radiative transfer may be important. Furthermore, the ionization state can be out of thermodynamical equilibrium. These factors might largely complicate the computations and 
these extra physics will be interesting follow-up work.

The density floor is set to maintain code stability when calling the equation of state. However, the presence of such a floor also sets a minimum mass which can be resolved in our simulation. To estimate that, we consider the outermost grid cells used in our simulation, located at a radius $\sim1.5 \times 10^{14}$ cm. Using the density floor, the minimum mass traceable by this grid is $\sim 6 \times 10^{-4} M_\odot$. The corresponding masses for inner grids are smaller as the grid volume scales as $\sim r^2$. This means the ejected mass ranging from $\sim 10^{-2} - 6~M_{\odot}$ in our models is likely well resolved by our code. For ejecta masses $\lesssim 0.01~M_{\odot}$, a suitable extension to even lower density below $ 10^{-11}$ g cm$^{-3}$ will be necessary. However, extending to lower densities will also require additional physical considerations, such as the validity of thermodynamic equilibrium and the inclusion of radiative processes.

Fourth, our two-dimensional simulations might underestimate small-scale turbulent features. Hydrodynamical instabilities, such
as the Kelvin-Helmholtz instabilities and Rayleigh-Taylor instabilities, can extend to the Kolmogorov length scale, where the viscous nature of the fluid becomes prominent. 
In fact, for a three-dimensional model, the development of Rayleigh-Taylor
instability can be different because the extra dimension allows
an extra degree of freedom for the development of the smaller features. 
This phenomenon is frequently seen in large-scale eddy modeling
where two-dimensional and three-dimensional turbulent energy cascades 
have different energy flow direction. However, the 
large-scale features reported here should be similar in the two types of simulations.

Fifth, we do not consider the effects of recombination and radiative transfer in these simulations. In our progenitor model, the outer $\approx 0.1 \, M_\odot$ has an optical depth $\tau \lesssim 10$, hence the assumption of high optical depth in those layers is not very good. While the approximation of  radiation being in equilibrium with matter is well maintained in the rest of the star, the true ejecta mass of our less energetic models (with $M_{\rm ej} \lesssim 0.1 \, M_\odot$) may be reduced if radiation can leak out before their outer layers are accelerated toward the escape speed.

The effects of recombination can also potentially affect the expansion of the envelope. Our equation of state does not include recombination energy, and the ionization temperature of $^{4}$He ($\sim$29000 K) and $^{1}$H ($\sim$10000 K) overlaps with the temperature range used in the simulation, meaning that partial recombination should take place during its expansion. Because of the envelope's low binding energy, this energy contributes substantially to the energy budget, and it may drive greater envelope expansion and mass loss than predicted by these simulations. On the other hand, recombination reduces the envelope opacity, allowing thermal energy to leak out, as described above. Resolving this competition between energy deposition by recombination, and energy loss by radiation, will be needed for better estimates of the mass loss and pre-SN density profile. 

Sixth, in the simulations we have made approximations for 
the initial conditions and energy deposition. In fact, how the 
exact energy deposition takes place in such a thin mass slice,
and how the initial convective structure can be coupled to 
a hydrostatic model, should be further explored. 

To clarify how these limitations affect the final results, we provide in the 
appendices more comparison tests. In Appendix A, we study how the initial density perturbation affects the later evolution without energy deposition. In Appendix B, we examine how the choice of density perturbation affects the later evolution. In Appendix C, we further compare how our two-dimensional energy deposition differs from the previous reported one-dimensional energy deposition model in \cite{Fuller2017}. In Appendix D, we study how our choice of smearing mass for the  energy deposition affects the morphology and dynamics of the ejecta. In Appendix E, we provide further details on our coordinate system. The above tests aim for clarifying the physics necessary for a comprehensive parameter survey for our future work.

\subsection{Conclusion}

In this work, we study the multidimensional hydrodynamical response of a red supergiant envelope heated by wave energy transport from nuclear burning in the core. We extract the stellar envelope from a 15 $M_{\odot}$ stellar model, and we inject energy similar to that expected from the wave luminosity and duration of the core oxygen-burning phase. To understand sensitivity to the uncertain wave luminosity and duration, we treat the heat deposition rate and its duration as model parameters. Unsurprisingly, we find that heat deposition in excess of the envelope's binding energy can drive large amounts of mass loss, while heat deposition lower than this threshold has a much smaller impact.

We also observe two classes of behaviors that depend on the wave heating rate. For a lower wave energy flux, the wave energy causes gradual expansion of the stellar envelope without strong mass loss, and the expansion is mostly spherical. With a high wave energy flux, the wave energy can amplify initial density asphericities and trigger uneven heating. This results in more asymmetric motion in the envelope where the Rayleigh Taylor instability arises, smoothing radial density inversions that occur in one-dimensional models. Larger heating rates also drive stronger shocks, expelling material above the original photosphere of the star. These low-density tails above the star's photosphere may have important effects on the early light curves of Type II-P SNe. Such events would also likely lead to observable pre-SN outbursts.

Comparison between our two-dimensional models and spherically symmetric models show only modest differences, mostly arising from Rayleigh-Taylor instabilities that smooth out density inversions created by wave heating. Hence, one-dimensional models are likely to yield reasonably good estimates of the unbound mass and the photometric signals produced by pre-SN outbursts in red supergiants. However, multi-dimensional models or one-dimensional models incorporating effects of Rayleigh-Taylor instability \citep{duffell:16} should be used for estimates of the pre-SN density profile.

We further discuss how the results of this work can be applied to stars of different mass. More massive stars with more vigorous convection during oxygen burning can potentially result in large enough heating rates to launch shocks which trigger observable outbursts and mass loss from the stellar envelope before the star collapse. Less massive stars with lower wave heating rates are unlikely to suffer such outbursts, except in cases where degenerate burning flashes, convective shell mergers, or a different core structure permit larger wave fluxes into the envelope. The corresponding optical signal, and the implications for SNe light curves will be interesting follow-up work. 

\section{Acknowledgments}

We thank Paul Duffell for useful insight regarding the simulations and results. S.C.L. acknowledges support from grants HST-AR-15021.001-A and 80NSSC18K1017. JF is thankful for support through an Innovator Grant from The Rose Hills Foundation, and the Sloan Foundation through grant FG-2018-10515. We thank Frank X. Timmes for his open-source subroutines for the Helmholtz equation of state and the template for the 7-isotope nuclear reaction network. We also thank the developers of the stellar evolution code MESA for their efforts in making the code public. 

\software{MESA (v8118); \citep{Paxton2011, Paxton2013, Paxton2015,Paxton2017}}

\appendix

\section{Initial Model}
\label{sec:static}

In building the initial model, we have used the hydrostatic model computed
by the MESA code as the input. We map the one-dimensional data, containing
the temperature and chemical composition as a function of the mass coordinate
and radial distance, onto a two-dimensional polar grid by solving again
the hydrostatic equilibrium equation for the density profile. 
Here we check how the initial model varies with time when there
is no energy deposition. This helps us understand whether the inner mass cut and 
the mapping satisfies the hydrostatic
equilibrium. It also serves as a reference for how much the 
initial aspherical seed contributes to the initial motion.

\begin{figure}
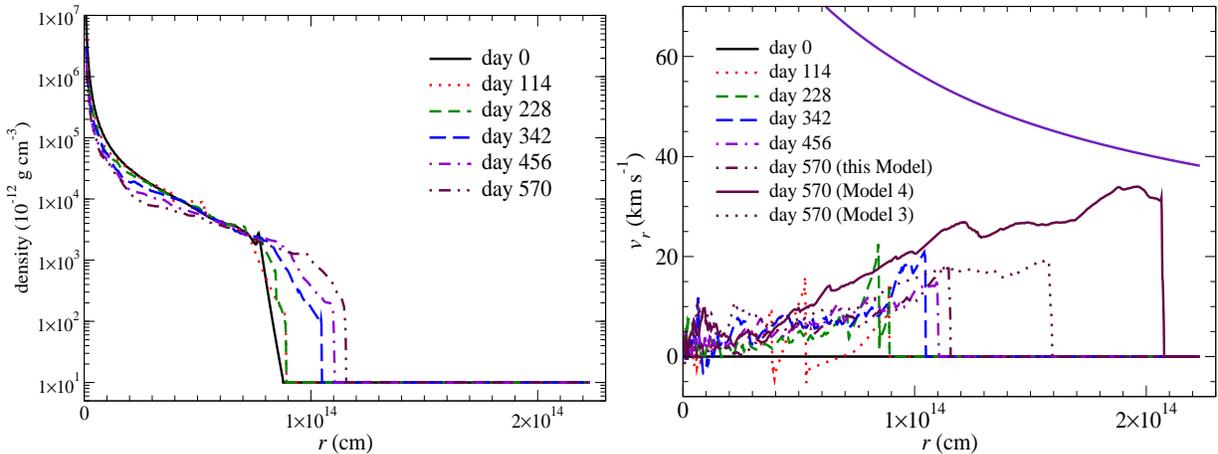

\centering
\includegraphics*[width=8cm,height=6cm]{rho_static_hydro_plot.eps}
\includegraphics*[width=8cm,height=6cm]{velr_static_hydro_plot.eps}
\caption{
{\bf Left:} The angle-averaged density of the test model without energy deposition.
{\bf Right:} Same as left panel but for the radial velocity.
For comparison, we also included velocity profiles of Models 3 and 4
at the same time points. The escape velocity (purple solid line) is included for reference. 
}
\label{fig:rho_static_time_plot}
\end{figure}

In Figure \ref{fig:rho_static_time_plot} we plot in the left and right 
panels the density and radial velocity profiles of the test model. The initial
model is exactly the same as those used in the main text, but without
any energy deposition during the simulation. From the density profiles, we can see that the star gradually relaxes to a new equilbrium that is slightly different from the initial conditions. The star mildly expands, indicated by the mild drop of the density at $2 - 7 \times 10^{13}$ cm and the outward motion of the surface. 
By the end of the simulation, the sharp cliff in density at the photosphere,
originally at $9 \times 10^{13}$ cm, is partially smoothed out. 

From the velocity profile, we can see the initial conditions do create some internal motion, leading to radial velocities of $\sim 20$ km s$^{-1}$, which is much lower than the escape velocity at $\sim 60$ km s$^{-1}$.
However, by comparing with the models from the main text, we can see the motion is much smaller than those with even low amounts of energy deposition, such as Model 3. Those models have a  final radius of $\sim 1.5$ -- $2$ times of the initial radius, while the test model only expands by $\sim 20\%$.
Hence, we can be confident that the majority of the response in the heated models is due to the deposited energy, as opposed to artifacts of the initial conditions

\section{Effects of Initial Aspherical Seeds}
\label{appendix:seed}

In the main text, we have described how we added density perturbations as a source of initial asphericity for enhancing the later development of aspherical motion and instabilities. We used a somewhat arbitrary value of $1 \%$. However, the actual density perturbations due to convection would certainly be different.

\begin{figure}
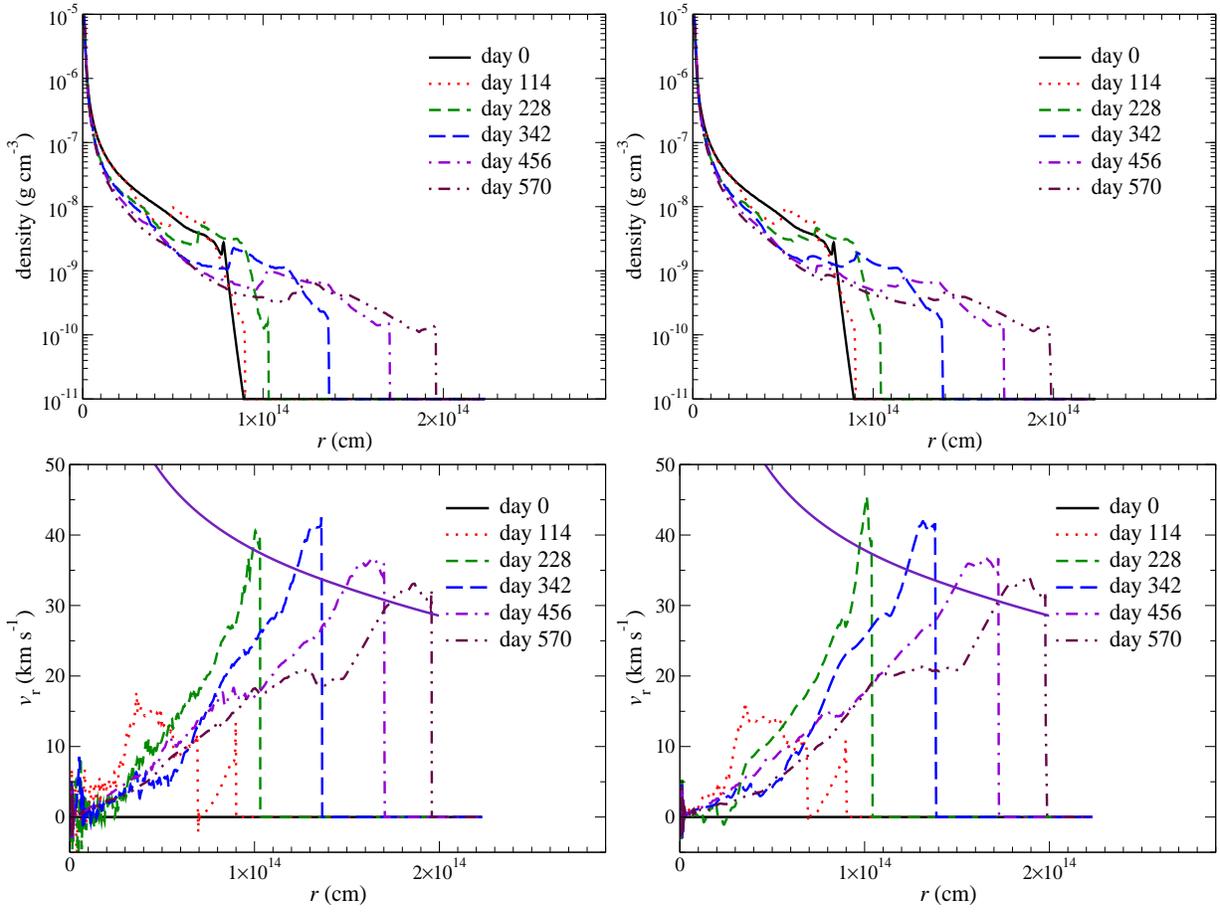

\centering
\includegraphics*[width=8cm,height=6cm]{rho_Model4b_plot.eps}
\includegraphics*[width=8cm,height=6cm]{rho_Model4a_plot.eps}
\includegraphics*[width=8cm,height=6cm]{velr_Model4b_plot.eps}
\includegraphics*[width=8cm,height=6cm]{velr_Model4a_plot.eps}
\caption{
{\bf Top left:} The angle-averaged density profiles of the test model based on Model 4 but with initial asphericity seed of magnitude $\sim 0.1 \%$. 
{\bf Top right:} Same as top left panel but for the model with initial
asphericity seed of magnitude $\sim 10 \%$. 
{\bf Bottom left:} Same as top left panel but for the radial velocity profile.
{\bf Bottom right:} Same as top right panel but for the radial velocity profiles.  
}
\label{fig:rho_seed_time_plot}
\end{figure}

Here we 
examine how the initial density perturbation affect the energy deposition.
We do not perturb the velocity fields because we find that the vector nature
of the velocity is more difficult to construct a good profile provides a 
good conservation of mass throughout the star, where density spans
more than 6 orders of magnitude. In Figure \ref{fig:rho_seed_time_plot} we plot the density, and velocity for two models identical to model 4, but with initial density perturbations smaller and larger by an order of magnitude. In both models, the behavior is very similar to that of Model 4.
The similarity of the density profiles give 
us indicates that most of our
results are insensitive to the details of the asphericity and density perturbations that would be present in a real star. 

The velocity profiles of these models further indicates insensitivity to the initial density perturbations. Comparing the profiles of two models at the same time slice, the differences are only $5-10 \%$. The similarity of the velocity profiles also ensures that our main results are robust against our choice of initial conditions.


\begin{figure}
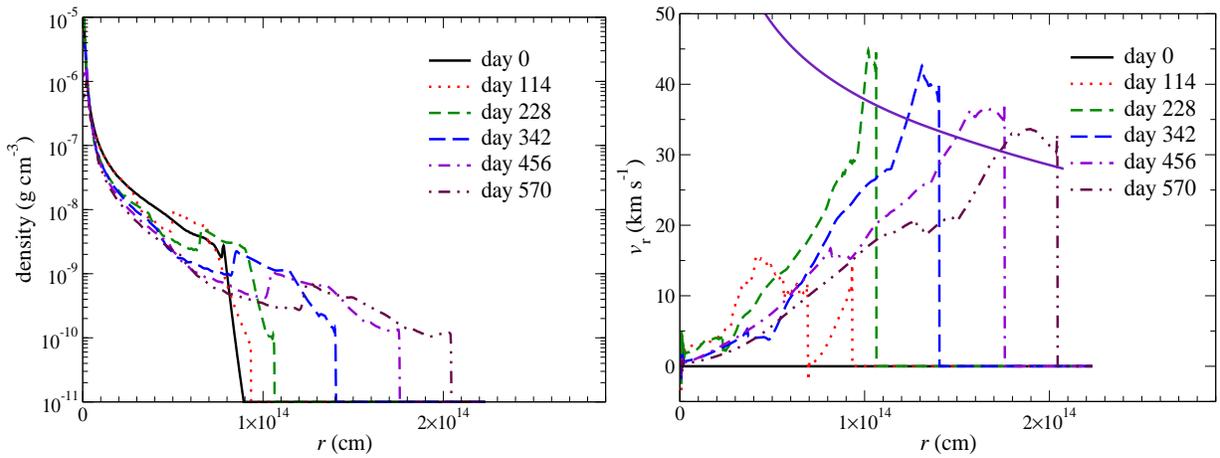

\centering
\includegraphics*[width=8cm,height=6cm]{rho_Model4s_plot.eps}
\includegraphics*[width=8cm,height=6cm]{velr_Model4s_plot.eps}
\caption{
{\bf Left:} The angle-averaged density profiles of Model 4s, which is identical to Model 4 but without density perturbations added to the initial conditions. 
{\bf Right:} Same as the left panel but for the velocity profiles.
}
\label{fig:rho_sph_time_plot}
\end{figure}


For comparison, we also compute a model without any initial density perturbations, shown in Figure \ref{fig:rho_sph_time_plot}.
Interestingly, this model shows a stiffer structure than
our previous models, more akin to the one-dimensional models of \cite{Fuller2017}. The density inversion is slightly larger and with a steeper boundary, likely because the initial spherical symmetry results in Rayleigh-Taylor instabilities that take longer to grow and smooth it out.
We also computed the Fourier asymmetry components, $A_n$ and they are a few orders of magnitude lower than our models presented in the main text. The velocity profiles in the right panel of Figure \ref{fig:rho_sph_time_plot} are very similar to the models with density perturbations. We conclude that density perturbations should be added to capture the realistic growth of Rayleigh-Taylor instabilities, but that the magnitude of the perturbations do not strongly affect our results.
Furthermore, we remark that in this simulation, the spherical symmetry is strictly preserved that the polar velocity field remains zero throughout the simulation. This ensures us that the simulation is equivalent to the  one-dimensional simulation counterpart that the pressure at a constant-radius slice is always balanced such that no spurious polar flow can be triggered when the model starts with spherical symmetry.

\section{Effects of Energy Deposition Mass}
\label{sec:Edepmass}

\begin{figure}
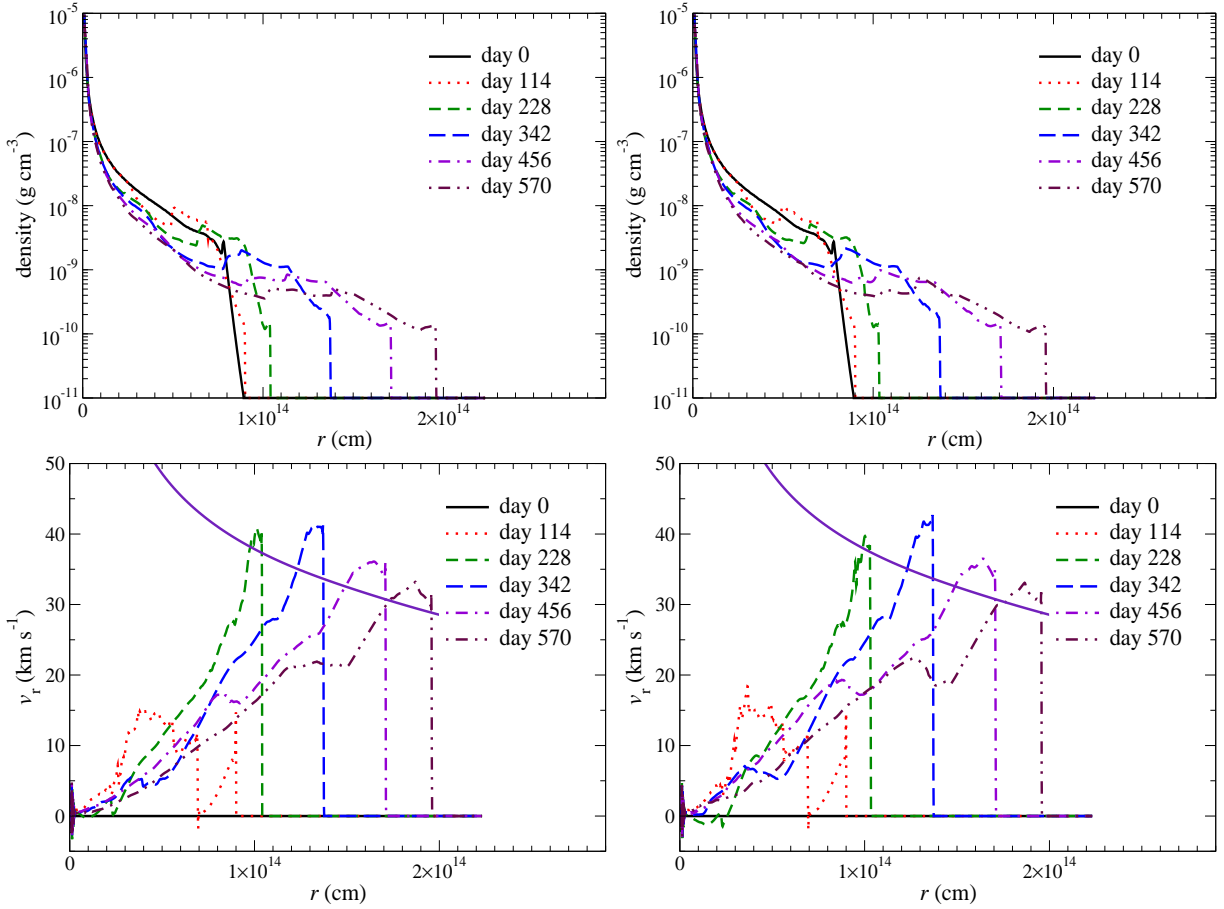

\centering
\includegraphics*[width=8cm,height=6cm]{rho_Model4c_plot.eps}
\includegraphics*[width=8cm,height=6cm]{rho_Model4d_plot.eps}
\includegraphics*[width=8cm,height=6cm]{velr_Model4c_plot.eps}
\includegraphics*[width=8cm,height=6cm]{velr_Model4d_plot.eps}
\caption{
{\bf Top left:} The angle-averaged density profiles of Model 4c, with energy deposition mass reduced to $0.05 ~M_{\odot}$. 
{\bf Top right:} Same as top left panel but for Model 4d with the energy deposition mass increased to 0.2 $M_{\odot}$. 
{\bf Bottom left:} Same as top left panel but for the radial velocity profiles.
{\bf Bottom right:} Same as top right panel but for the radial velocity profiles.  
}
\label{fig:rho_mdep_time_plot}
\end{figure}

Here we compare how the choice of smoothing the energy deposition over mass affects our results.
When we calculate how the waves deposit the energy, we always smear the full energy deposition into a mass of $M_{\rm dep}$ because in extreme cases the deposition sphere can recede towards the inner boundary. This poses a challenging modeling scenario as the sharp drop of the density and sound speed profile can cause all the energy to be deposited in a single grid cell, such that the energy deposition is not well resolved. In fact, for the higher $L_{\rm dep}$ case ($10^7~ L_{\odot}$), the deposition length scale can become much smaller than a grid cell, so it is numerically impossible to resolve the exact distribution of energy. To alleviate the problem, we define $M_{\rm dep}$ such that the energy deposition is always smeared into a few grid cells, such that our results are less sensitive to resolution. 

To test the effect of this parameter on our results, we perform two extra models, Model 4c and Model 4d, which have $M_{\rm dep} = 0.05$ and $0.2 ~M_{\odot}$ respectively. In Figure \ref{fig:rho_mdep_time_plot}, we plot the density and velocity profiles for the two models. Despite the fact that the deposition mass differs by 4 times between the two models, we can see that there are almost no observable differences, except minor changes in the peak velocity and the details of the density gradient. This provides evidence that the smearing procedure provides a good enough description for the energy deposition process in a resolvable manner.

\section{Global Comparison of Test Models}
\label{sec:global_test}

\begin{figure}
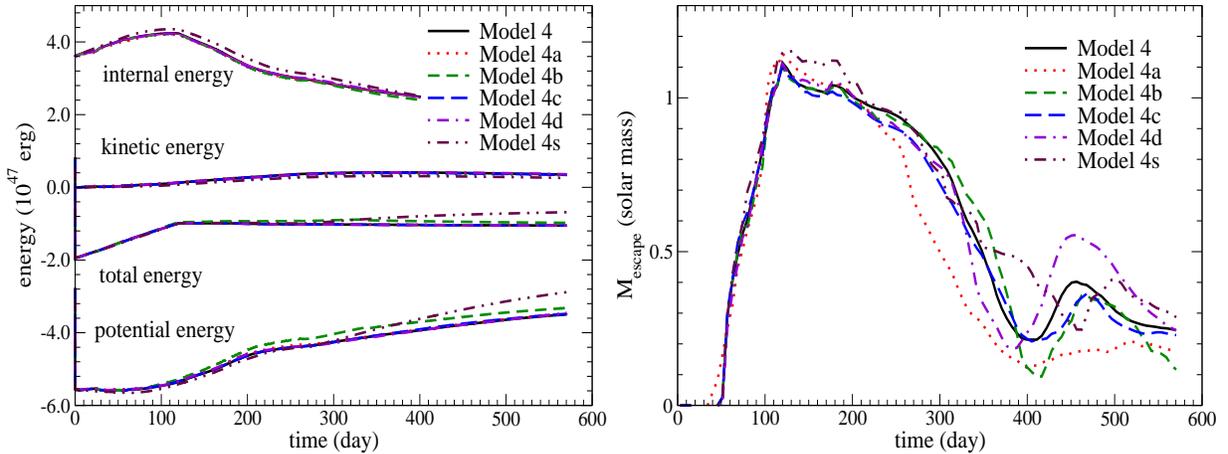

\centering
\includegraphics*[width=8cm,height=6cm]{energy_time_test_plot.eps}
\includegraphics*[width=8cm,height=6cm]{ejecta_mass_test_plot.eps}
\caption{
{\bf Left:} The total energy, kinetic, gravitational and internal
energy of Model 4, 4a, 4b, 4c, 4d
and 4s. 
{\bf Right:} Same as left
panel but for the ejecta mass.
}
\label{fig:global_time_plot}
\end{figure}

Next, we compare the global energetics and the ejecta mass for the test cases examined in the appendix. They serve as a general diagnostic as how sensitive each controllable parameter is to our final result. In Figure \ref{fig:global_time_plot}, we plot the total energy and its components for the test models, including Model 4 as a reference. We also plot the ejecta mass as a function of time. We can see that the choice of $M_{\rm dep}$ or aspherical seed size does not significantly change the energetics of the star. Model 4s shows a slightly larger expansion as described in above, thus it has a slightly higher potential energy due to more energy inflow from the atmosphere. 

In regards to the ejecta mass, Figure \ref{fig:global_time_plot} shows that all the models are qualitatively the same, suggesting that the ejecta mass is not sensitive to the parameters presented above. All models show their peak at Day 120 at $\sim 1.1~M_{\odot}$ and then the unbound mass gradually decreases until day 400 at 0.1--0.4 $M_{\odot}$. At that point the models begin to deviate somewhat, but the final ejecta mass is always in the range 0.1--0.3 $M_{\odot}$.

\section{Notes on the Geometry}


Our wedge simulations can be considered to be pieces of the star's equatorial or meridional plane. We use periodic boundary conditions in $\theta$ so that matter leaving one $\theta$ boundary enters the other $\theta$ boundary.
For the computations of mass and energy, we define the volume elements ${\rm vol}(r,\theta) = 8 r^2 dr d\theta$ for a quadrant, automatically giving the volume of a sphere when integrated in both $r$ and $\theta$. In this choice, the 2D grid cells are intersections of 3D volume elements with the orbital plane, with each 3D volume element wrapping from the positive $z$-axis to the negative $z$-axis at constant radius. In the literature, axisymmetric models in the meridional plane are more frequently used for two-dimensional simulations. In the absence of rotational symmetry breaking forces, our geometry is physically identical to the meridional plane. We choose this geometry over a meridional plane geometry (in which case the volume elements would be $4 \pi r^2 dr \sin \theta d \theta$) to give equal weight to all latitudes $\theta$. Additionally, we choose periodic boundary conditions because a closed boundary in the polar direction can strongly suppress the polar flow in the simulation.

\begin{figure}
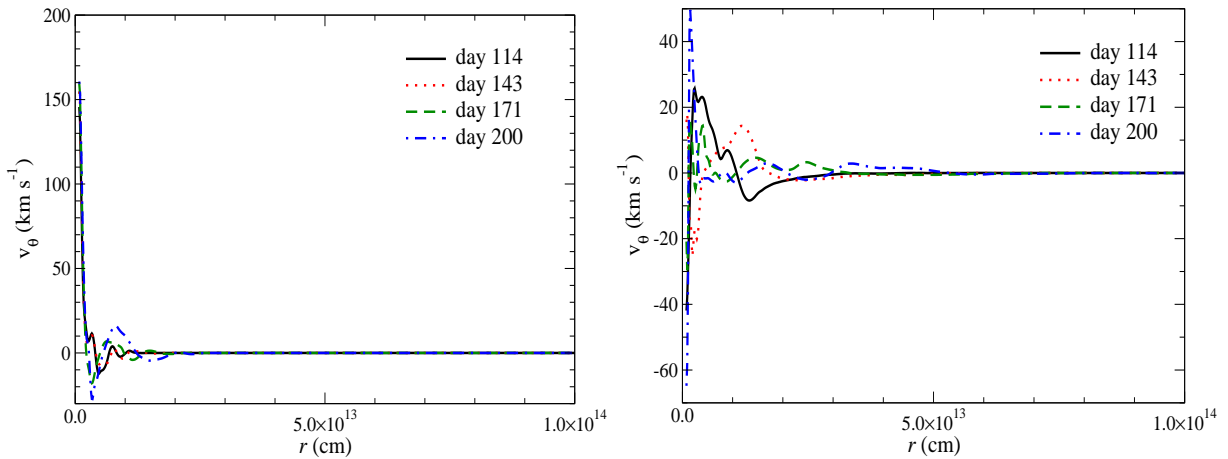

\centering
\includegraphics*[width=8cm,height=6cm]{velz_Model4_plot2.eps}
\includegraphics*[width=8cm,height=6cm]{velz_Model4cb_plot.eps}
\caption{
{\bf Left:} The angular velocity of Model 4 at selected time frames.
{\bf Right:} Same as left panel but for Model 4 with reflective angular boundaries.
}
\label{fig:vel_boundary_plot}
\end{figure}

In the main text, we noted that the energy deposition may generate a substantial horizontal flow in the inner layers of the star. To examine the choice of boundary conditions on this flow, we recompute Model 4 with reflective angular boundary conditions, explicitly suppressing any flow across those boundaries. We show in Figure \ref{fig:vel_boundary_plot} the angular velocities in both cases.
In the model with periodic boundaries, a high-velocity flow has developed and remains substantial from day 114. On the other hand, the model with reflective boundaries shows lower velocities near the inner boundary, but similar velocities everywhere else.
While the heating may drive mean flows in the inner layers, the angular motion appears turbulent in the outer part of the domain. The two models indicate that
the dynamics of the outer part of the star are insensitive to the boundary conditions, but the possibility of rotational flows in the inner part of the star should be examined in future work.

\bibliographystyle{apj}
\pagestyle{plain}
\bibliography{biblio}

\end{document}